\documentclass[12pt]{article}
\usepackage[active]{srcltx}
\usepackage{epsfig}
\usepackage{graphicx}
\usepackage{cite}
\newcommand{\eref}[1]{(\ref{#1})}

\newcommand{\na}{\mbox{\boldmath $\nabla$}}

\newcommand{\E}{\mbox{\boldmath $E$}}

\newcommand{\Hb}{\mbox{\boldmath $H$}}
\newcommand{\av}{\mbox{\boldmath $a$}}

\newcommand{\cE}{\mbox{\boldmath$\cal E$}}
\newcommand{\cH}{\mbox{\boldmath$\cal H$}}
\newcommand{\Rho}{\mbox{\boldmath $\hat{\cal R}$}}
\newcommand{\Ta}{\mbox{\boldmath $\hat{\cal T}$}}

\renewcommand{\t}{\mbox{\boldmath $t$}}
\newcommand{\D}{\mbox{\boldmath $D$}}

\newcommand{\cc}{\mbox{\boldmath $c$}}

\newcommand{\eps}{\mbox{\boldmath $\varepsilon$}}
\newcommand{\ka}{\mbox{\boldmath $\kappa$}}
\newcommand{\ps}{\mbox{\boldmath $\psi$}}
\newcommand{\Ps}{\mbox{\boldmath $\Psi$}}

\newcommand{\e}{\mbox{\boldmath $e$}}

\newcommand{\n}{\mbox{\boldmath $n$}}

\newcommand{\rr}{\mbox{\boldmath $r$}}

\newcommand{\J}{\mbox{\boldmath $J$}}

\newcommand{\bb}{\mbox{\boldmath $b$}}
\newcommand{\B}{\mbox{\boldmath $B$}}

\newcommand{\kk}{\mbox{\boldmath $k$}}

\newcommand{\lv}{\mbox{\boldmath$l$}}
\newcommand{\hI}{\mbox{\boldmath$\hat{I}$}}

\newcommand{\hE}{\mbox{\boldmath $\hat{E}$}}
\newcommand{\hT}{\mbox{\boldmath $\hat{T}$}}
\newcommand{\hR}{\mbox{\boldmath $\hat{R}$}}

\newcommand\fr{\displaystyle\frac}

\newcommand\ola{\overleftarrow}
\newcommand\ora{\overrightarrow}
\newcommand{\htts}{\mbox{\boldmath$\hat{t}\kern1pt$}}

\renewcommand{\tan}{{\rm\ tg}}
\newcommand\lt{\left}
\newcommand\rt{\right}

\newcommand{\skk}{\mbox{\scriptsize\boldmath$k$}}
\newcommand{\srr}{\mbox{\scriptsize\boldmath$r$}}

\newcommand{\qm}{quantum mechanics}

\begin{document}
\begin{center}
Ignatovich F.V.$^1$, Ignatovich V.K.$^{2*}$,

$^1$ Lumetrics inc, Rochester, N.Y., USA

$^2$FLNP JINR, Dubna, Russia
\bigskip

{\Large\bf Bulk and surface plane electromagnetic waves in
anisotropic media}
\end{center}

\begin{abstract}
A new analytical approach to description of electromagnetic waves
in nonmagnetic anisotropic media is presented. Amplitudes of their
reflection and refraction at interfaces and also reflection and
transmission of plane parallel plates are derived. Beam splitting
at reflection, and creation of surface waves at the interfaces are
studied. A simple laboratory demonstration of the beam splitting
is proposed. D'yakonov surface waves, their description and
observation are discussed.
\end{abstract}

\hfill\parbox{9.3cm}{\it ---Reading the manuscript, this reviewer
decided that the authors must be new to electromagnetics.
Otherwise, they would not have wasted their time reinventing the
wheel. Everything that could be known about propagation in an
uniaxial material has been known since at least 1841. So there is
no sense wasting time there.}

\hfill{Referee of Opt.Technol.}

\section{Introduction}

Description of electromagnetic waves in homogeneous anisotropic
media did not change since Fresnel times more than 160 years ago.
Here the first time since then we present a new approach, which
makes physics here very simple and transparent. We will not waist
time describing the standard approach, which can be found with
only slight variation in all the textbooks on electrodynamics or
optics [1-13] containing chapters on anisotropic media. Instead we
directly start with our approach.

An anisotropic medium is characterized by some direction, called
axis which we will denote by a unit vector $\av$. In a plane
electromagnetic wave
\begin{equation}\label{3}
\cE\exp(i\kk\rr-i\omega t)
\end{equation}
propagating in such a medium at an arbitrary direction with
respect to $\av$ the wave vector $k(\omega)$ and polarization
vector $\cE$ depend on angle $\theta$ between $\av$ and direction
of propagation $\ka=\kk/k$.

The main feature of our approach (similar to the one used for
elastic waves~\cite{igl}) is a special representation of the
dielectric permittivity tensor $\eps$. It was proven by
Fedorov~\cite{fed} that in a uniaxial anisotropic medium the
tensor $\eps$ can be represented as
\begin{equation}\label{1}
\eps_{ij}=\epsilon_1\delta_{ij}+\epsilon'a_ia_j,
\end{equation}
where $\epsilon_1$ is isotropic part, and anisotropy is
characterized by the unit vector $\av$ with components $a_i$ and
by anisotropy parameter $\epsilon'$.

In the next section we find $\cE$ and $k(\omega)$ in anisotropic
media with one and two mutually orthogonal axes. In the third
section we discuss reflection of these waves from interfaces
between anisotropic and isotropic media, study beam splitting at
reflection, conditions for creation of surface waves, and propose
a device to demonstrate the beam splitting at reflections in a
laboratory. In section 4 we calculate reflection and transmission
for plane parallel anisotropic plates. In 5-th section we
calculate speed of the D'yakonov surface waves~\cite{dya}, which
are an analog of Rayleigh elastic waves on a free surface of
elastic media. We correct some defect of the derivation of these
waves in~\cite{dya} and propose an experiment to generate and to
observe the D'yakonov waves. In 6-th section we summarize our
results. In sections 3-5 we limit ourselves only to uniaxial media
and only explain the ideas. An approach to biaxial media and most
heavy mathematics are shifted to appendices.

\section{Plane waves in anisotropic media}
\label{se1}

\hfill\parbox{6.3cm}{\it ---I don't get the goal of this paper}

\hfill{Referee of Am.J.Phys.}

The wave equation is derived from Maxwell equations, which in the
absence of currents and charges are
\begin{equation}\label{3a}
-[\na \times \E(\rr,t)] = \frac{\partial}{c\partial t}\B(\rr,t),
\quad [\na \times \Hb(\rr,t)] = \frac{\partial}{c\partial
t}\D(\rr,t), \quad \na\B=0, \quad \na\D=0,
\end{equation}
where
\begin{equation}\label{3b}
\B=\mu H, \qquad \D=\eps\E,
\end{equation}
and $\mu$, $\eps$ are magnetic and dielectric permittivities. In
the following we take $\mu=1$, and then the equations \eref{3a}
are simplified to
\begin{equation}\label{3c}
-[\na \times \E(\rr,t)] = \frac{\partial}{c\partial t}\Hb(\rr,t),
\quad [\na \times \Hb(\rr,t)] = \frac{\partial}{c\partial
t}\eps\E(\rr,t), \quad \na\Hb=0, \quad \na\eps\E=0.
\end{equation}
Differentiation of the second equation of \eref{3c} over time and
after that substitution of the first equation leads to the wave
equation
\begin{equation}\label{2}
-[\na\times[\na \times \E(\rr,t)]]= \fr{\partial^2}{c^2\partial
t^2}\eps\E(\rr,t).
\end{equation}

\subsection{Electric vector of the wave}

Substitution of the plane wave \eref{3} into this equation reduces
\eref{2} to
\begin{equation}\label{4}
k^2\cE-\kk(\kk\cdot\cE)=k_0^2\eps\cE,
\end{equation}
where $k_0=\omega/c$. Eq. \eref{4} is valid only in homogeneous
media. If we have an interface between two homogeneous media we
have two different equations of the type \eref{1} in them, and
matching of waves in two media is performed via boundary
conditions, which follow not from the wave equation itself, like
in \qm, but from Maxwell equations.

In uniaxial anisotropic media we use tensor $\eps$ in the form
\eref{1}. Therefore for a plane wave \eref{3} we have
\begin{equation}\label{a4}
\eps\cE=\epsilon_1\cE+\epsilon'\av(\av\cdot\cE),
\end{equation}
and the last equation in \eref{3c} leads to
\begin{equation}\label{6}
\epsilon_1(\kk\cdot\cE)+\epsilon'(\kk\cdot\av)(\av\cdot\cE)=0.
\end{equation}
Substitution of \eref{a4} into \eref{4} gives
\begin{equation}\label{b4}
k^2\cE-\kk(\kk\cdot\cE)-k_0^2\eps\cE\equiv(k^2-k_0^2\epsilon_1)\cE-\kk(\kk\cdot\cE)-k_0^2\epsilon'\av(\av\cdot\cE)=0.
\end{equation}
To find $\cE$ we need to solve \eref{b4} with account of \eref{6}.

The 3-dimensional vector $\cE$ can be represented by coordinates
in some basis. If $\kk$ is not parallel to $\av$, we can use as a
basis three independent vectors $\av$, $\ka=\kk/k$ and
\begin{equation}\label{e1}
\e_1=[\av\times\ka].
\end{equation}
In this basis $\cE$ looks
\begin{equation}\label{8}
\cE=\alpha\av+\beta\ka+\gamma\e_1
\end{equation}
with coordinates $\alpha$, $\beta$ and $\gamma$, which are not
independent, because of Eq. \eref{6}.

Substitution of \eref{8} into \eref{6} gives
\begin{equation}\label{9}
\epsilon_1[k\beta+(\kk\cdot\av)\alpha]+\epsilon'(\kk\cdot\av)[\alpha+\beta(\ka\cdot\av)]=0,
\end{equation}
 from which it follows that
\begin{equation}\label{10}
\beta=-\fr{(\ka\cdot\av)(1+\eta)} {1+\eta(\ka\cdot\av)^2}\alpha,
\end{equation}
where $\eta=\epsilon'/\epsilon_1$. Substitution of \eref{10} into
\eref{8} gives
\begin{equation}\label{11}
\cE=\alpha\lt(\av-\ka\fr{(\ka\cdot\av)(1+\eta)}
{1+\eta(\ka\cdot\av)^2}\rt)+\gamma\e_1=\alpha\e_2+\gamma\e_1,
\end{equation}
which shows that $\cE$ lies in a plane of two independent vectors
$\e_1=[\av\times\ka]$ and the orthogonal to it
\begin{equation}\label{12}
\e_2=\av-\ka(\ka\cdot\av)\fr{1+\eta} {1+\eta(\ka\cdot\av)^2}\equiv
\av-\ka(\ka\cdot\av)\epsilon_2(\theta)/\epsilon_1,
\end{equation}
where $\cos\theta=\ka\cdot\av$ and we introduced anisotropic
dielectric permittivity
\begin{equation}\label{12a}
\epsilon_2(\theta)=\epsilon_1\fr{1+\eta} {1+\eta\cos^2\theta}.
\end{equation}

To find coordinates $\alpha$ and $\beta$ we substitute \eref{11}
into \eref{b4} and multiply it by $\e_1$. As a result we obtain
\begin{equation}\label{12b}
(k^2-k_0^2\epsilon_1)\gamma\e_1^2=0.
\end{equation}
It shows that if $\gamma\ne0$, then \eref{12b} can be satisfied
only when
\begin{equation}\label{13}
k^2=k_0^2\epsilon_1.
\end{equation}
Multiplying \eref{b4} by $\av$ and taking into account that
\begin{equation}\label{a13}
\av\cdot\e_2=\fr{1-(\ka\cdot\av)^2}{1+\eta(\ka\cdot\av)^2}, \qquad
\ka\cdot\e_2=-\eta(\ka\cdot\av)\fr{1-(\ka\cdot\av)^2}{1+\eta(\ka\cdot\av)^2}=-\eta(\ka\cdot\av)(\av\cdot\e_2),
\end{equation}
we obtain
\begin{equation}\label{12c}
\lt(k^2-k_0^2\epsilon_2(\theta)\rt)\alpha(\av\cdot\e_2)=0.
\end{equation}
Therefore, if $\alpha\ne0$, and $\av\ne\ka$, Eq. \eref{12c} can be
satisfied only, when
\begin{equation}\label{12d}
k^2=k_0^2\epsilon_2(\theta),
\end{equation}
where $\epsilon_2(\theta)$ is given in \eref{12a}. Since the
length of $k$ is different for two polarization vectors, therefore
a single plain wave can exist only with a single polarization
along either $\e_2$ or $\e_1$.

We will call ``transverse'' the mode with polarization
$\cE_1=\e_1$, and ``mixed'' the mode with polarization along
$\cE_2=\e_2$. The mixed mode according to \eref{a13} contains a
longitudinal component along the wave vector $\ka$. We think that
such a nomenclature is better than common names: ``ordinary'' for
$\cE_1=\e_1$, and ``extraordinary'' for $\cE_2=\e_2$, because our
names point to physical features of these waves.

\subsection{Magnetic fields}

Every electromagnetic wave besides electric contains also magnetic
field. From the equation $\na\cdot\Hb=0$, which is equivalent to
$\kk\cdot\Hb=0$, it follows that the field $\Hb$ is orthogonal to
$\kk$. It is also orthogonal to $\cE$, which follows from the
first equation of \eref{3c}. After substitution of Eq. \eref{3}
into \eref{3c} and the field $\Hb$ in the plane wave form
\begin{equation}\label{3a2}
\Hb(\rr,t)=\cH\exp(i\kk\cdot\rr-i\omega t),
\end{equation}
with polarization vector $\cH$ we obtain
\begin{equation}\label{3a1}
\cH=\fr k{k_0}[\ka\times\cE].
\end{equation}
For transverse and mixed modes in uniaxial media, respectively, we
therefore obtain
\begin{equation}\label{16a}
\cH_1=\fr
k{k_0}[\ka\times\e_1]=\fr{k}{k_0}[\ka\times[\av\times\ka]],\qquad
\cH_2=\fr{k}{k_0}[\ka\times\e_2]=\fr{k}{k_0}[\ka\times\av],
\end{equation}
and the total plain wave field looks
\begin{equation}\label{16a1}
\Ps(\rr,t)=\ps_j\exp(i\kk_j\cdot\rr-i\omega t),
\end{equation}
where $\ps_j=\cE_j+\cH_j$, and $j$ denotes mode 1 or 2. In
isotropic media we also can choose, say $\cE=[\av\times\ka]$ and
$\cH=[\ka\times[\av\times\ka]]$. However, there $\av$ can have
arbitrary direction, therefore the couple of orthogonal vectors
$\cE$ and $\cH$ can be rotated any angle around the wave vector
$\kk$.

\section{Reflection from an interface between uniaxial and\\ isotropic medium}

\hfill\parbox{9.3cm}{{\it ---I spent a few separate sittings
reading this paper but cannot convince myself to sit with pencil
and paper and follow the mathematics presented here, since I know
I can open Jackson,... or Griffith and get really all I would need
to know about the interaction of E\&M waves with the surface of a
dielectric medium--in a more
streamlined, concise and easy to follow manner.}\\\small Referee of Am.J.Phys.\\
{\bf A note} Anisotropic media are not considered
in~\cite{Jack,grif}.} \medskip

Imagine that our space is split into two half spaces. The part at
$z<0$ is a uniaxial anisotropic medium, and the part at $z>0$ is
vacuum with $\epsilon_1=1$, $\eta=0$. We have two different wave
equations in these parts, and waves go from the reign of one
equation into the reign of another one through the interface where
they must obey boundary conditions imposed by Maxwell equations.

Let's look for reflection of the two possible modes incident onto
the interface from within the anisotropic medium.

\subsection{Nonspecularity and mode transformation at the interface}

First we note that reflection of the mixed mode is not specular.
Indeed, since direction of $\kk$ after reflection changes,
therefore the angle $\theta$ between $\av$ and $\ka$ does also
change, and $k$, according to \eref{12d}, changes too. However the
component $\kk_\|$ parallel to the interface does not change, so
the change of $k$ means the change of the normal component
$k_\bot$, and this leads to nonspecularity of the reflection.

Let's calculate the change of $k_\bot$ for the incident mixed mode
with wave vector $\kk_{2r}$, where the index r means that the mode
2 propagates to the right toward the interface. For a given angle
$\theta$ between $\kk_{2r}$ and $\av$ we can write
\begin{equation}\label{17}
k_{2r\bot}=\sqrt{\fr{\epsilon_1k_0^2(1+\eta)}{1+
\eta\cos^2\theta}-k_\|^2},
\end{equation}
however the value of $k_{2r\bot}$ enters implicitly into
$\cos\theta$, so to find explicit dependence of $k_{2r\bot}$ on
$\av$ it is necessary to solve the equation
\begin{equation}\label{18}
k_\|^2+x^2+\eta(k_\|(\lv\cdot\av)+x(\n\cdot\av))^2=k_0^2\epsilon_1(1+\eta),
\end{equation}
where $x$ denotes $k_{2r\bot}$, $\n$ is a unit vector of normal,
directed toward isotropic medium, and $\lv$ is a unit vector along
$\kk_\|$, which together with $\n$ constitutes the plane of
incidence. Solution of this equation is
\begin{equation}\label{19}
x=\fr{-\eta
k_\|(\n\cdot\av)(\lv\cdot\av)+\sqrt{\epsilon_1k_0^2(1+\eta)(1+\eta(\n\cdot\av)^2)-
k_\|^2(1+\eta(\lv\cdot\av)^2+\eta(\n\cdot\av)^2)}}{1+\eta(\n\cdot\av)^2}.
\end{equation}
The sign chosen before square root provides the correct
asymptotics at $\eta=0$ equal to isotropic value
$\sqrt{\epsilon_1k_0^2- k_\|^2}$.

In general vector $\av$ is representable as
$\av=\alpha\n+\beta\lv+\gamma\t$, where $\t=[\n\lv]$ is a unit
vector perpendicular to the plane of incidence. The normal
component $k_{2r\bot}$ depends only on part of this vector
$\av'=\alpha\n+\beta\lv$, which lies in the incidence plane. If we
denote $\alpha=|\av'|\cos(\theta_a)$,
$\beta=|\av'|\sin(\theta_a)$, where $|\av'|$ is projection of
$\av$ on the incidence plane, and introduce new parameter
$\eta'=\eta|\av'|^2\le\eta$, then formula \eref{19} is simplified
to
\begin{equation}\label{19a}
k_{2r\bot}=\fr{-\eta'
k_\|\sin(2\theta_a)+2\sqrt{\epsilon_1k_0^2(1+\eta)[1+\eta'\cos^2(\theta_a)]-
k_\|^2(1+\eta')}}{2[1+\eta'\cos^2(\theta_a)]}.
\end{equation}

For the reflected mixed mode (mode 2, propagating to the left from
the interface) an equation similar to \eref{18} looks
\begin{equation}\label{20}
k_\|^2+x^2+\eta(k_\|(\lv\cdot\av)-x(\n\cdot\av))^2=k_0^2\epsilon_1(1+\eta),
\end{equation}
where $x=k_{2l\bot}$, and its solution is
\begin{equation}\label{21}
k_{2l\bot}=\fr{\eta'
k_\|\sin(2\theta_a)+2\sqrt{\epsilon_1k_0^2(1+\eta)[1+\eta'\cos^2(\theta_a)]-
k_\|^2(1+\eta')}}{2[1+\eta'\cos^2(\theta_a)]}.
\end{equation}
We see that the difference of the normal components of reflected
and incident waves of mixed modes $k_{2l\bot}-k_{2r\bot}$ is
\begin{equation}\label{22}
k_{2l\bot}-k_{2r\bot}=\fr{\eta'
k_\|\sin(2\theta_a)}{1+\eta'\cos^2(\theta_a)}.
\end{equation}
In the following we will present such differences in dimensionless
variables
\begin{equation}\label{a22}
\Delta_{22}\equiv\fr{k_{2l\bot}-k_{2r\bot}}{k_0\sqrt{\epsilon_1}}=\fr{\eta'
q\sin(2\theta_a)+2\sqrt{(1+\eta)[1+\eta'\cos^2(\theta_a)]-
q^2(1+\eta')}}{2[1+\eta'\cos^2(\theta_a)]},
\end{equation}
where $q^2=k_\|^2/k_0^2\epsilon_1$. The reflection angle depends
on orientation of anisotropy vector $\av$ and it can be both
larger than the specular one, when $\theta_a>0$, or smaller, when
$\theta_a<0$.

In the case of transverse incident mode the length $k=|\kk|$ of
the wave vector, according to \eref{13}, does not depend on
orientation of $\av$, therefore this wave is reflected specularly.
\begin{figure}[t!]
{\par\centering\resizebox*{8cm}{!}{\includegraphics{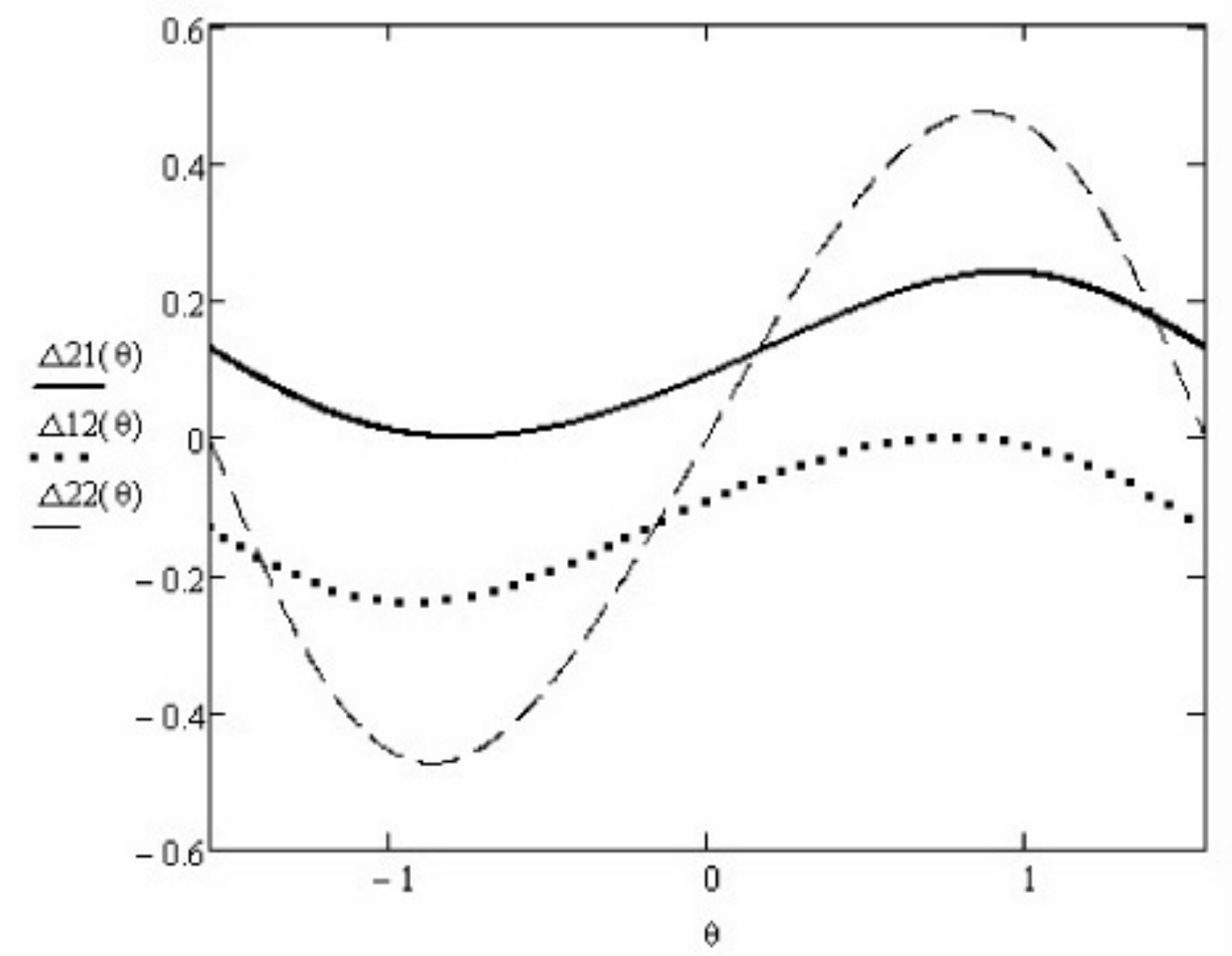}}\par}
\caption{Variation of change $\Delta$ of normal components for
reflected and incident waves in dependence of angle
$\theta=\theta_a$ of anisotropy vector with respect to normal
$\n$. Vector $\av$ is supposed to lie completely in the incidence
plane. The curves $\Delta ij$ represent dimensionless ratio
$\Delta_{ij}(\theta)$ given by \eref{a22}, \eref{22a} and
\eref{22b}. The curves were calculated for $\eta=\eta'=0.4$ and
$q=k_\|/k_0\sqrt{\epsilon_1}=0.7$.} \label{f1}
\end{figure}

Every incident mode after reflection creates another one, because
without another mode it is impossible to satisfy the boundary
conditions. Let's look what will be the normal component of the
wave vector of other mode. If the incident is the wave of mode 2,
reflected transverse mode (mode 1 propagating to the left, away
from the interface) will have
$k_{1l\bot}=\sqrt{\epsilon_1k_0^2-k_\|^2}$. Therefore according to
\eref{19a} the difference
$\Delta_{12}=(k_{1l\bot}-k_{2r\bot})/k_0\sqrt{\epsilon_1}$ is
\begin{equation}\label{22a}
\Delta_{12}=\sqrt{1-q^2}-\fr{-\eta'
q\sin(2\theta_a)+2\sqrt{(1+\eta)[1+\eta'\cos^2(\theta_a)]-
q^2(1+\eta')}}{2[1+\eta'\cos^2(\theta_a)]}.
\end{equation}
In the opposite case, when the incident mode is transverse one,
the reflected mixed mode will have $k_{2l\bot}$ shown in
\eref{21}. Therefore the difference
$\Delta_{21}=(k_{2l\bot}-k_{1r\bot})/k_0\sqrt{\epsilon_1}$ is
\begin{equation}\label{22b}
\Delta_{21}=\fr{\eta'
q\sin(2\theta_a)+2\sqrt{(1+\eta)[1+\eta'\cos^2(\theta_a)]-
q^2(1+\eta')}}{2[1+\eta'\cos^2(\theta_a)]}-\sqrt{1-q^2}.
\end{equation}
The changes of normal components with variation of $\theta_a$
according to \eref{a22}, \eref{22a} and \eref{22b} for some values
of dimensionless parameters $\eta$ and $q$ and vector $\av$ lying
completely in the incidence plane, are shown in Fig. \ref{f1}.
From this figure it is seen that the strongest deviation of
reflected wave from specular direction is observed for reflection
of mixed to mixed mode.
\begin{figure}[t!]
{\par\centering\resizebox*{6cm}{!}{\includegraphics{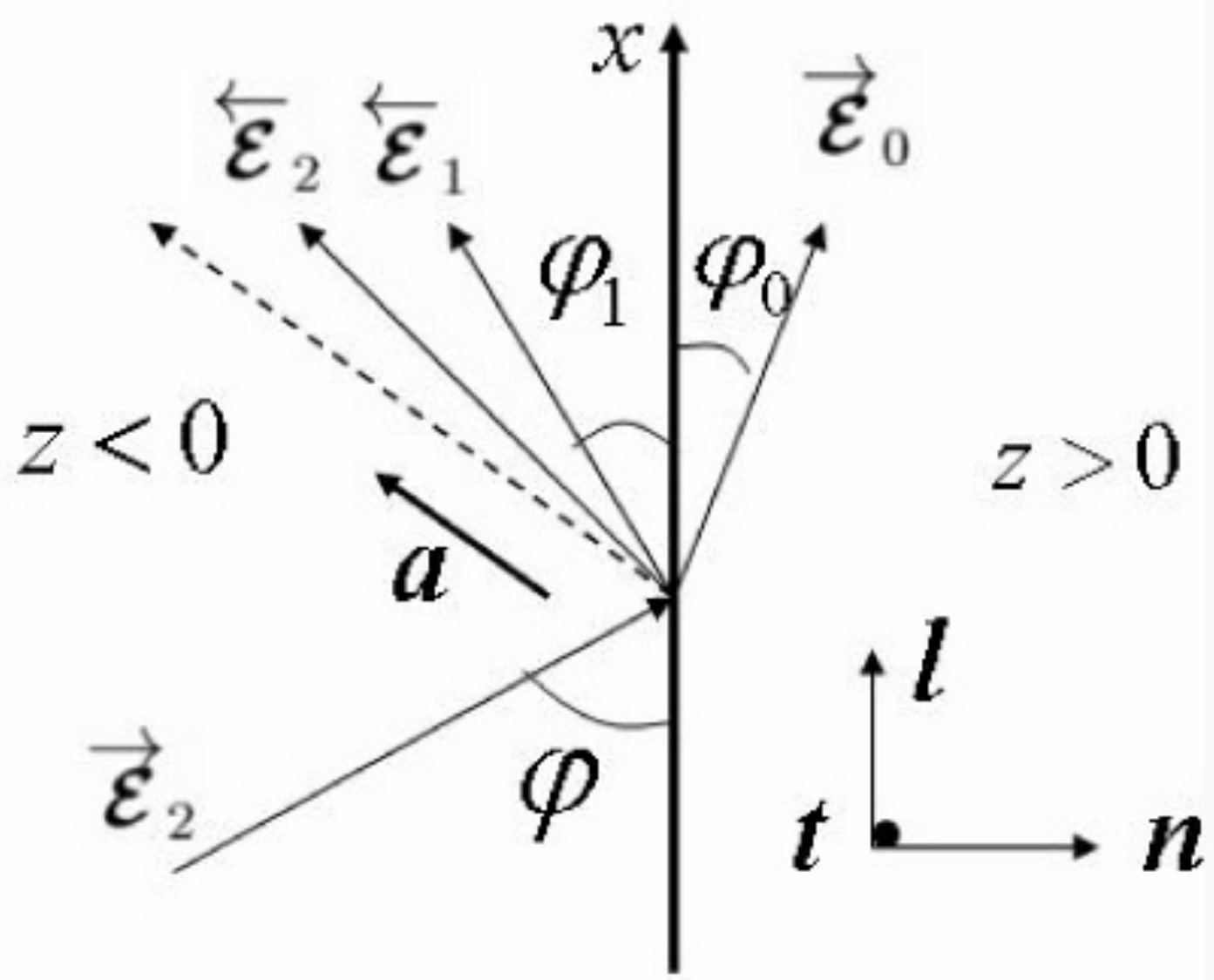}}\par}
\caption{Arrangement of wave vectors of all the modes created by
the incident wave of mode 2,  i.e. of polarization
$\protect\ora\cE_2$, when the anisotropy vector $\av$ has the
direction shown here. The grazing angle of the reflected mode 2,
$\protect\ola\cE_2$, is less than specular one (specular direction
is shown by the broken arrow), and the grazing angle $\varphi_1$
of the reflected mode 1, $\protect\ola\cE_1$, is even lower. The
grazing angle $\varphi_0$ of the transmitted wave
$\protect\ora\cE_0$ is even lower than $\varphi_1$. We can imagine
than at some critical value $\varphi=\varphi_{c1}$ the angle
$\varphi_0$ becomes zero. It means that at $\varphi<\varphi_{c1}$
transmitted wave becomes evanescent and all the incident energy is
totally reflected in the form of two modes. More over, there
exists a second critical angle $\varphi_{c2}$, when $\varphi_1=0$.
Below this angle at $\varphi<\varphi_{c2}$ the mode
$\protect\ola\cE_1$ also becomes evanescent. In this case all the
incident energy is totally reflected nonspecularly in the form of
the mode 2. At the same time the two evanescent waves
$\protect\ora\cE_0$ and $\protect\ola\cE_2$ combine into a surface
wave, propagating along the interface. The arrows over $\cE$ show
direction of waves propagation with respect to the interface. In
the figure there is also shown the basis which is used along the
paper. It consists of unit normal vector $\n$ along normal
($z$-axis), unit tangential vector $\lv$ ($x$-axis) which together
with $\n$ defines the incidence plane, and the vector $\t$
($y$-axis) looking toward the reader, which is normal to the
incidence plane.} \label{f2}
\end{figure}

Since reflection of mode 2 is in general nonspecular, it can
happen that the wave vectors of reflected and transmitted waves
will be arranged as shown in fig. \ref{f2}, and it follows that
there are two critical angles for $\varphi$. The first critical
angle $\varphi_{c1}$ ($q^2=1/\epsilon_1$) is the angle of total
reflection. The transmitted wave at it becomes evanescent. The
totally reflected field contains two modes. At the second critical
angle $\varphi_{c2}$, when $q$ is in the range
\begin{equation}\label{023}
1<q^2<\fr{(1+\eta)(1+\eta'cos^2(\theta_a))}{1+\eta'}.
\end{equation}
\begin{figure}[b!]
{\par\centering\resizebox*{8cm}{!}{\includegraphics{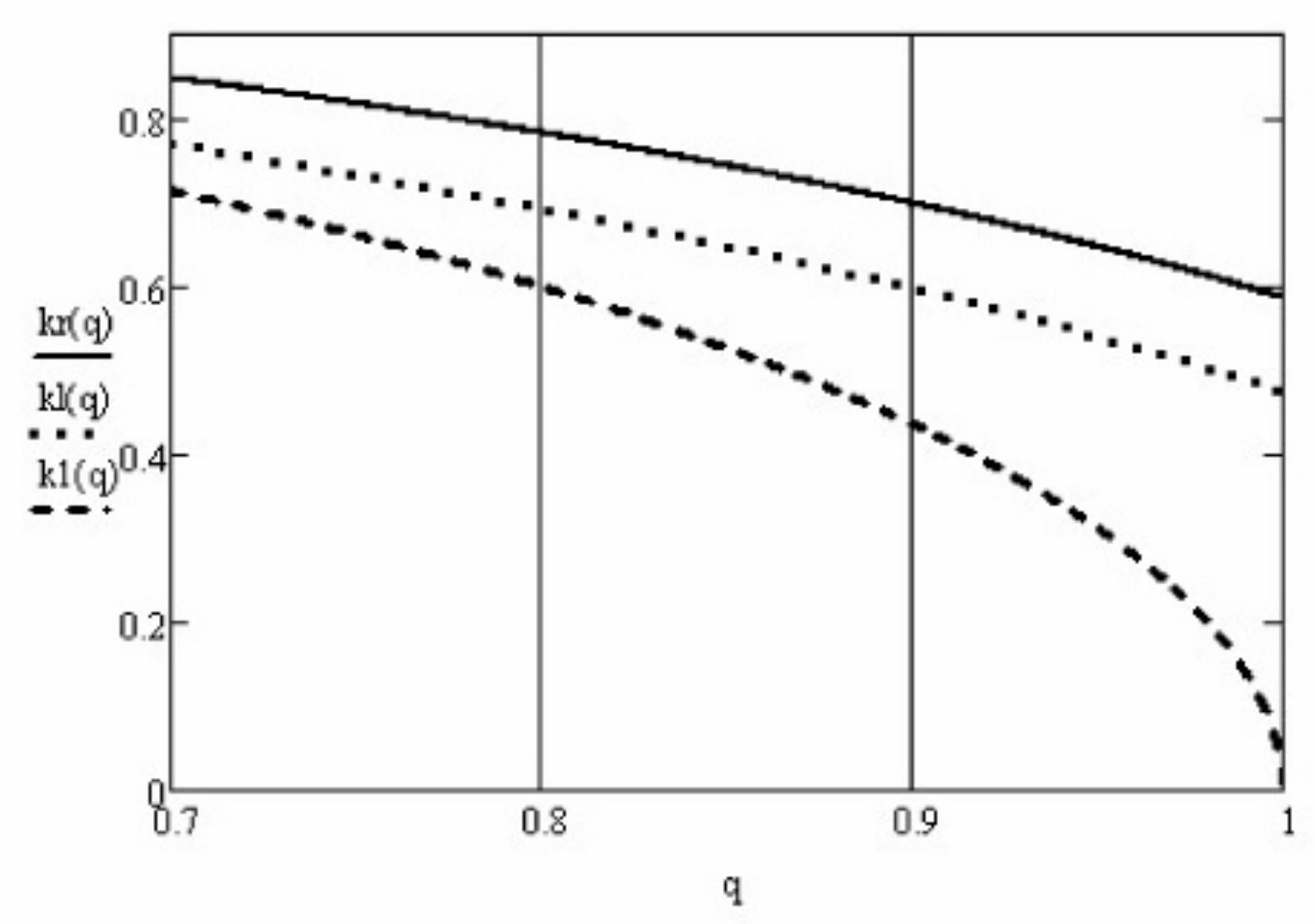}}\par}
\caption{Dependence of dimensionless normal components of incident
and reflected waves on $q=k\cos\varphi/k_0\sqrt{\epsilon_1}$. The
solid curve corresponds to the incident wave moving to the right
$kr(q)=k_{2r\bot}/k_0\sqrt{\epsilon_1}$. The dotted curve
corresponds to the reflected wave of mode 2 moving to the left
$kl(q)=k_{2l\bot}/k_0\sqrt{\epsilon_1}$. And the broken curve
corresponds to the reflected wave of mode 1 moving to the left
$k1(q)=k_{1l\bot}/k_0\sqrt{\epsilon_1}$. It is seen that at $q>1$
the mode 1 ceases to propagate. Its normal component
$k1(q)=\sqrt{1-q^2}=-i\sqrt{q^2-1}$ becomes imaginary, therefore
the reflected mode 1 becomes an evanescent wave. Together with
transmitted wave, which becomes evanescent at $q^2=1/\epsilon_1$,
the mode 1 constitute the surface electromagnetic wave.}
\label{f3}
\end{figure}
the reflected mode 1 also becomes evanescent. Together with
evanescent transmitted wave the mode 1 constitutes a surface wave,
propagating along the interface. In that case we have nonspecular
total reflection of the single mode $\cE_2$ and the surface wave
tied to it.

In figure \ref{f3} it is shown how do the normal components of
wave vectors change with increase of $q$, which is equivalent to
decrease of $\varphi$. For $\epsilon_1=1.6$ the first critical
angle corresponds to $q\approx0.8$. The second critical angle
corresponds to $q=1$.

\subsection{Demonstration of the beam splitting with the help of a
birefringent cone}

\hfill\parbox{9.3cm}{{\it --- In order to ensure that our journal
continues to publish articles that cater for our broad readership,
every paper submitted must meet our stringent editorial criteria.
We believe that your article does not meet these criteria, so it
has been withdrawn from consideration.}\\ \small Editor of
J.Phys.A Optics}

\begin{figure}[h!]
{\par\centering\resizebox*{8cm}{!}{\includegraphics{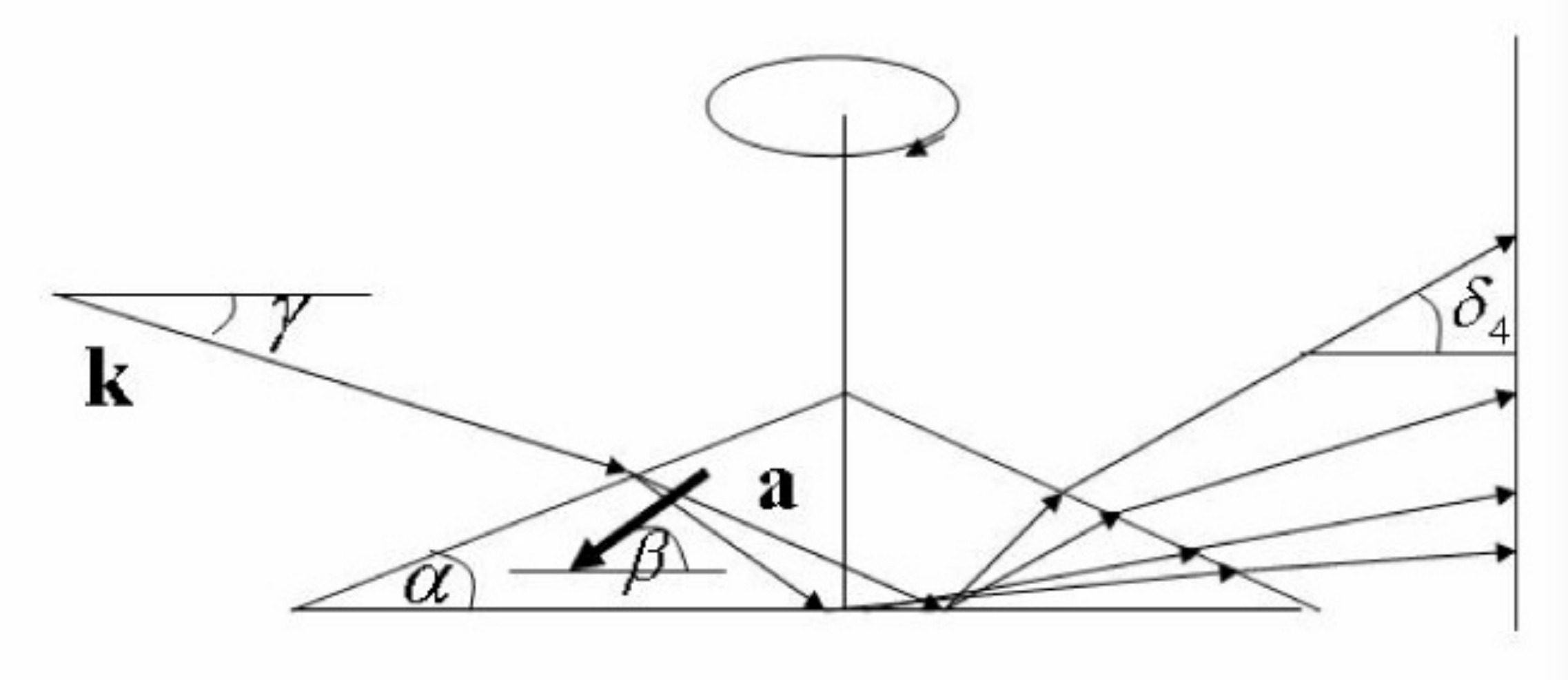}}\par}
\caption{\label{f} Demonstration of the beam splitting of light in
a birefringent cone. Bright spots on a vertical screen change
their position and brightness when the cone is rotated around
vertical axis.}
\end{figure}
Beam splitting at interfaces of an anisotropic medium can be
spectacularly demonstrated with the help of birefringent cone as
shown in fig.\ref{f}. In the geometrical optics approximation a
narrow incident beam of light after refraction on the side surface
of the cone is split into two rays of two different modes 1 and 2.
Both modes are further split into two components at reflection
from basement of the cone. Four resulting beams after refraction
at the side surface go out of the cone and produce on a vertical
screen four bright spots. Their positions and brightness depend on
direction of the anisotropy axis inside the cone and vary with the
cone rotation.

The direct numerical calculations for parameters $\epsilon_1=1.6$,
$\eta=0.8$, $\av$ in the figure plane, and $\sin\alpha=0.5$,
$\sin\beta=0.3$ ¨ $\sin\gamma=0.5$ show that outgoing beams from
below to top have directions characterized by $\tan\delta_1=0.2$,
$\tan\delta_2=0.4$, $\tan\delta_3=0.6$ and $\tan\delta_4=0.7$
respectively.

\section{Calculation of refraction at interfaces and scattering on plates}

\hfill\parbox{9.3cm}{{\it --- Too much of a mathematical exercise
and too little physics for Physica Scripta.}\\ \small Referee of
Phys.Scripta}\medskip

To calculate reflection and transmission of a plane parallel
anisotropic plate, placed in isotropic (for instance, vacuum)
medium, it is necessary to know reflections and refractions at
interfaces from inside and outside anisotropic medium, which is
obtained by imposing boundary conditions stemmed from Maxwell
equations. Knowledge of everything at interfaces permits to write
directly reflection and transmission of the plate by the method
which is explained in~\cite{igu} and will be shortly described
below.

\subsection{Reflection and refraction
from inside of the anisotropic medium}

The wave function in the full space is
\begin{equation}\label{23}
\Psi(\rr)=\Theta(z<0)\lt(e^{i\ora{\skk}_{j}\cdot\srr}\ora{\psi}_{j}+\sum\limits_{j'=1,2}
e^{i\ola{\skk}_{j'}\cdot\srr}\ola{\psi}_{j'}\ora\rho_{j'j}\rt)+
\Theta(z>0)e^{i\skk_0\cdot\srr}\Big(\psi_{e}\ora\tau_{ej}+\psi_{m}\ora\tau_{mj}\Big),
\end{equation}
where $\Theta$ is a step function equal to unity,when inequality
in its argument is satisfied, and to zero in opposite case, half
space $z<0$ is occupied by anisotropic medium, and the half space
$z>0$ is vacuum, $\psi=\cE+\cH$, arrows show direction of waves
propagation, $\ora{\psi}_{j}$ denotes the incident wave of mode
$j$ ($j=1,2$), $\ola{\psi}_{j'}$ ($l=1,2$) denotes reflected wave
of mode $j'$, $\ora{\kk}_{j}=(\kk_\|,k_{jr\bot})$,
$\ola{\kk}_{j'}=(\kk_\|,-k_{j'l\bot})$,
$\kk_0=(\kk_\|,\sqrt{k_0^2-k_\|^2})$, $\psi_{e,m}$,
$\ora\tau_{e,mj}$ are the refracted fields and refraction
amplitudes of TE- and TM-modes respectively for the incident
$j$-mode. To find reflection $\ora\rho$ and refraction $\ora\tau$
amplitudes (the arrow over them shows the direction of propagation
of the incident wave toward the interface), we need to impose on
\eref{23} the following boundary conditions.

\subsection{General equations from boundary conditions}

Every incident wave field can be decomposed at the interface into
TE- and TM-modes. In TE-mode electric field is perpendicular to
the incidence plane, $\cE\propto\t$, therefore contribution of
$j$-th mode into TE-mode is $(\cE_j\cdot\t)$. In TM-mode magnetic
field is perpendicular to the incidence plane, $\cH\propto\t$,
therefore contribution of $j$-th mode into TM-mode is
$(\cH_j\cdot\t)$. For refracted field in TE-mode we accept
$\ora\cE_e=\t$, $\ora\cH_e=[\ka_0\times\t]$, and for refracted
field in TM-mode we accept $\ora\cH_m=\t$,
$\ora\cE_m=-[\ka_0\times\t]$.

\subsubsection{TE-boundary conditions}

In TE-mode for incident j-mode we have the following three
equations from boundary conditions:
\begin{enumerate}
    \item continuity of electric field
\begin{equation}\label{a23}
(\t\cdot\ora\cE_j)+(\t\cdot\ola\cE_1)\ora\rho_{1j}+(\t\cdot\ola\cE_2)\ora\rho_{2j}=(\t\cdot\ora\cE_e)\ora\tau_{ej},
\end{equation}
    \item continuity of magnetic field parallel to the interface
\begin{equation}\label{a23a}
(\lv\cdot\ora\cH_j)+(\lv\cdot\ola\cH_1)\ora\rho_{1j}+(\lv\cdot\ola\cH_2)\ora\rho_{2j}=(\lv\cdot[\ka_0\times\t])\ora\tau_{ej}\equiv
-\kappa_{0\bot}\ora\tau_{ej},
\end{equation}
    \item and continuity of the normal component of magnetic induction,
    which for $\mu=1$ looks
\begin{equation}\label{b23a}
(\n\cdot\ora\cH_j)+(\n\cdot\ola\cH_1)\ora\rho_{1j}+(\n\cdot\ola\cH_2)\ora\rho_{2j}=(\n\cdot[\ka_0\times\t])\ora\tau_{ej}\equiv
\kappa_{0\|}\tau_{ej}.
\end{equation}
\end{enumerate}
The last eq. \eref{b23a} is, in fact, not needed, because it
coincides with \eref{a23}.

\subsubsection{TM-boundary conditions}

In TM-mode we have the equations
\begin{enumerate}
    \item continuity of magnetic field
\begin{equation}\label{c23}
(\t\cdot\ora\cH_j)+(\t\cdot\ola\cH_1)\ora\rho_{1j}+(\t\cdot\ola\cH_2)\ora\rho_{2j}=\ora\tau_{mj},
\end{equation}
\item continuity of electric field parallel to the interface
\begin{equation}\label{d23a}
(\lv\cdot\ora\cE_j)+(\lv\cdot\ola\cE_1)\ora\rho_{1j}+(\lv\cdot\ola\cE_2)\ora\rho_{2j}=-(\lv\cdot[\ka_0\times\t])\ora\tau_{mj}
\equiv\kappa_{0\bot}\ora\tau_{mj},
\end{equation}
\item and continuity of the normal component of field $\D$
    \begin{equation}\label{e23a}
(\n\cdot\eps\ora\cE_j)+(\n\cdot\eps\ola\cE_1)\ora\rho_{1j}+(\n\cdot\eps\ola\cE_2)\ora\rho_{2j}=
(\n\cdot[\ka_0\times\t])\ora\tau_{mj}\equiv\kappa_{0\|}\ora\tau_{mj}.
\end{equation}
\end{enumerate}
Again we can neglect Eq. \eref{e23a}, because it coincides with
\eref{c23}. In the following we will not show third equations like
\eref{b23a} and \eref{e23a}, because they are useless. Solution of
all the equation is presented in Appendix B. Here for simplicity
we limit ourselves only to a particular case of normal incidence
of the waves.

\hfill\parbox{9.3cm}{{\it --- The authors should not have allowed
themselves to use the phrase "for simplicity".}\\\small Referee of
Am.J.Phys.}

\subsubsection{A particular case of normal incidence}

In the case of normal incidence reflection and refraction are
especially simple, because there is no splitting at reflection. We
define geometry by three basis vectors $\n$, $\lv$ and $\t$, where
$\n$ denotes normal directed along $z$ axis toward vacuum, while
vectors $\lv$ and $\t$ lie in the interface and define $x$ and $y$
axes respectively. The anisotropy vector $\av$ is supposed to lie
in $(x,z)$ plane at angle $\theta$ with respect to $\n$.

A plane wave propagating along $\n$ ($\ka=\kk/k=\n$) can have only
two types of polarizations. It can be of a transverse mode with
$\cE_1=\e_1\equiv\t$ and $\cH_1=-n_1\lv$, where
$n_1=\sqrt{\epsilon_1}$ (see \eref{1}), or it can be of mixed mode
with $\cE_2=\e_2$ (\eref{12}), and
$\cH_2=n_2(\theta)[\n\times\av]$, where
$n_2(\theta)=\sqrt{\epsilon_2(\theta)}$ (see \eref{12a}).

Since there are no splitting at normal incidence the boundary
conditions are simplified. For the mode 1 \eref{a23} and
\eref{a23a} are reduced to
\begin{equation}\label{1a23}
[1+\ora\rho_{11}]=\ora\tau_{e1},\qquad
n_1[1-\ora\rho_{11}]=\ora\tau_{e1},
\end{equation}
and transmitted wave has field polarization $\cE_{e1}=\t$,
$\cE_{e1}=\lv$ identical to that of the incident field.

Solution of \eref{1a23} is
\begin{equation}\label{3a23}
\ora\rho_{11}=\fr{n_1-1}{n_1+1},\qquad
\ora\tau_{e1}=\fr{2n_1}{n_1+1},\qquad n_1=\sqrt{\epsilon_1}.
\end{equation}
With these formulas we can immediately find reflection and
transmission of a plane parallel plate of thickness $D$ for an
incident from vacuum electromagnetic wave with polarization
$\cE_{1e}=\t$:
\begin{equation}\label{a1a23}
R_1=-\ora\rho_{11}\fr{1-\exp(2ik_1D)}{1-\ora\rho^2_{11}\exp(2ik_1D)},\qquad
T_1=\exp(ik_1D)\fr{1-\ora\rho^2_{11}}{1-\ora\rho^2_{11}\exp(2ik_1D)},
\end{equation}
where $k_1=k_0n_1$. We see that the incident wave with linear
polarization $\cE_{e1}=\t$ parallel to that of mode 1 inside the
plate does not change polarization after transmission through the
plate.

Now let's apply boundary conditions to the mode $\cE_2$ incident
normally on to the interface from inside the plate. Now the
boundary conditions are reduced to
\begin{equation}\label{11a23}
(\lv\cdot\av)[1+\ora\rho_{22}(\theta)]=\ora\tau_{e2},\qquad
k_2(\theta)(\lv\cdot\av)[1-\ora\rho_{22}(\theta)]=k_0\ora\tau_{e2},
\end{equation}
where factor $(\lv\cdot\av)$ appears because of projection of the
polarization vector onto the interface. Therefore, since
$k_2=k_0n_2(\theta)$, and $n_2(\theta)=\sqrt{\epsilon_2(\theta)}$
solution of \eref{11a23} is
\begin{equation}\label{2a23}
\ora\rho_{22}(\theta)=\fr{n_2(\theta)-1}{n_2(\theta)+1},\qquad
\ora\tau_{e2}(\theta)=\fr{2n_2(\theta)(\lv\cdot\av)}{n_2(\theta)+1}.
\end{equation}

From symmetry consideration we can immediately find reflection and
transmission amplitudes for outside incident waves with unit
polarization along $\lv$:
\begin{equation}\label{4a23}
\ola\rho_{2e}(\theta)=\fr{1-n_2(\theta)}{1+n_2(\theta)},\qquad
\ola\tau_{2e}(\theta)=\fr2{(\lv\cdot\av)(1+n_2(\theta))},
\end{equation}
and therefore we can immediately find reflection and transmission
of a plane parallel plate of thickness $D$ for an incident from
vacuum electromagnetic wave with polarization $\cE_{2e}=\lv$:
\begin{equation}\label{b1a23}
R_2(\theta)=-\ora\rho_{22}(\theta)\fr{1-\exp(2ik_2(\theta)D)}{1-\ora\rho^2_{22}(\theta)\exp(2ik_2(\theta)D)},$$
$$T_2(\theta)=\exp(ik_2(\theta)D)\fr{1-\ora\rho^2_{22}(\theta)}{1-\ora\rho^2_{22}(\theta)\exp(2ik_2(\theta)D)},
\end{equation}
where $k_2(\theta)=k_0n_2(\theta)$. We see that the incident wave
with linear polarization $\cE_{2e}=\lv$, which lies in the plane
$(\n,\av)$, does not change polarization after transmission
through the plate.

Now let's consider transmission through the plate of a plane wave
$\exp(ik_0z-i\omega_0t)\cE_e$ with intermediate polarization:
$\cE_e=\alpha\t+\beta\lv$, where $|\alpha|^2+|\beta|^2=1$. The
transmitted electrical part of the wave will be
\begin{equation}\label{623}
\E_t(z,t)=\exp(ik_0(z-D)-i\omega_0t)\lt[\alpha T_1\t+\beta
T_2(\theta)\lv\rt].
\end{equation}
Since $\alpha T_1=|\alpha T_1|\exp(i\gamma_1)$, and $\beta
T_2(\theta)=|\beta T_2(\theta)|\exp(i\gamma_2(\theta))$, the real
part of the wave \eref{623} at some point $z$ chosen for
convenience so that $k_0(z-D)+\gamma_1=2\pi N$ with an integer
$N$, looks
\begin{equation}\label{523}
\rm{Re}(\E_t(z,t))=|\alpha T_1|\cos(\omega_0t)\t+|\beta
T_2(\theta)|\sin(\omega_0t-\varphi)\lv,
\end{equation}
where $\varphi=\gamma_{2}(\theta)-\gamma_1-\pi/2$. So, in this
case the transmitted field has elliptical polarization.

\subsection{Reflection and refraction
from outside an anisotropic medium}

Let's consider the case, when the half space at $z<0$ is vacuum,
and that at $z>0$ is an anisotropic medium. The incident wave goes
from the left in the vacuum. The wave function in the full space
now looks
\begin{equation}\label{2a3}
\Psi(\rr)=\Theta(z<0)\Big(e^{i\ora\skk_{0}\srr}\ora\psi_j+e^{i\ola\skk_{0}\srr}\sum_{j'=e,m}\ola\psi_{j'}\ora\rho_{j'j}\Big)+
\Theta(z>0)\lt(e^{i\ora\skk_{1}\srr}\ora\psi_{1}\ora\tau_{1j}+e^{i\ora\skk_{2}\srr}\ora\psi_{2}\ora\tau_{2j}\rt),
\end{equation}
where $j,j'$ denote $e$ or $m$ for TE- and TM-modes respectively,
the term $\exp(i\ora\kk_{0}\rr)\ora\ps_{j}$ with the wave vector
$\ora\kk_{0}=(\kk_\|,k_{0\bot}=\sqrt{k_0^2-k_\|^2})$ describes the
plain wave incident on the interface from vacuum. In TE-mode
factor $\ora\ps_{e}=\ora\cE_{e}+\ora\cH_{e}$ contains
$\ora\cE_{e}=\t$ and $\ora\cH_{e}=[\ora\ka_0\t]$. In TM-mode
factor $\ora\ps_{m}=\ora\cE_{m}+\ora\cH_{m}$ contains
$\ora\cH_{m}=\t$ and $\ora\cE_{m}=-[\ora\ka_0\t]$.

The reflected wave has the wave vector
$\ola\kk_{0}=(\kk_\|,-k_{0\bot})$, and fields $\ola\cE_{e}=\t$,
$\ola\cH_{e}=[\ola\ka_0\t]$, $\ola\cH_{m}=\t$, and
$\ola\cE_{m}=-[\ola\ka_0\t]$. The refracted field contains two
wave modes with wave vectors $\ora\kk_{1}=(\kk_\|,k_{1\bot})$,
$\ora\kk_{2}=(\kk_\|,k_{2r\bot})$ and electric fields
$\ora\cE_{1}=\e_1=[\av\ora\ka_{1}]$ and
$\ora\cE_{2}=\e_2=\av-\ora\ka_{2}(\av\ora\ka_{2})\epsilon_2(\ora\theta_2)/\epsilon_1$.
Here $\ka=\kk/k$, $k_{1\bot}=\sqrt{\epsilon_1k_0^2-k_\|^2}$, and
$k_{2r\bot}$ is given by \eref{19a}. For incident TE-mode
reflection $\rho_{ee}$, $\rho_{me}$ and refraction $\tau_{je}$
amplitudes ($j=1,2$) are found from boundary conditions
\begin{equation}\label{A23}
(\t\ora\cE_{1})\ora\tau_{1e}+(\t\ora\cE_{2})\ora\tau_{2e}=1+\ora\rho_{ee},
\end{equation}
\begin{equation}\label{A23a}
(\lv\ora\cH_{1})\ora\tau_{1e}+(\lv\ora\cH_{2})\ora\tau_{2e}=-\kappa_{0\bot}(1-\ora\rho_{ee}),
\end{equation}
\begin{equation}\label{C23}
(\t\ora\cH_{1})\ora\tau_{1e}+(\t\ora\cH_{2})\ora\tau_{2e}=\ora\rho_{me},
\end{equation}
\begin{equation}\label{d223a}
(\lv\ora\cE_{1})\ora\tau_{1e}+(\lv\ora\cE_{2})\ora\tau_{2e}=-\kappa_{0\bot}\ora\rho_{me}.
\end{equation}
Solution of these equations is elementary and is given in Appendix
C.

\subsection{Reflection and transmission amplitudes for a plane parallel plate
of thickness $L$}

Now, when we understand what happens at interfaces, we can
construct~\cite{igu} expressions for reflection, $\ora\hR(L)$, and
transmission, $\ora\hT(L)$, matrices for a whole anisotropic plane
parallel plate of some thickness $L$, when the state of the
incident wave is described by a general vector
$|\ora\xi_0\rangle$. To do that let's denote the state of field of
the modes $\e_1$ and $\e_2$ incident from inside the plate onto
the second interface at $z=L$ by unknown  2-dimensional vector
$|\ora x\rangle$ \eref{mx}. If we were able to find $|x\rangle$ we
could immediately write the state of transmitted field
\begin{equation}\label{ax}
\ora\hT(L)|\ora\xi_0\rangle=\ora{\Ta^{\,\prime}}|\ora x\rangle,
\end{equation}
and the state of the field, reflected from the whole plate
\begin{equation}\label{ksx}
\ora\hR(L)|\ora\xi_0\rangle=\ora\Rho|\xi_0\rangle+\ola{\Ta^{\,\prime}}\ola\hE(L)\ora{\Rho'}|\ora
x\rangle,
\end{equation}
where $\ola\hE(L)$, $\ora\hE(L)$ denote diagonal matrices
\begin{equation}\label{2x}
\ola\hE(L)=\left(%
\begin{array}{cc}
  \exp(i k_{1\bot}L) & 0 \\
 0 & \exp(ik_{2l\bot}L)\\
\end{array}%
\right),\qquad \ora\hE(L)=\left(%
\begin{array}{cc}
  \exp(ik_{1\bot}L) & 0 \\
 0 & \exp(ik_{2r\bot}L)\\
\end{array}%
\right).
\end{equation}
which describe propagation of two modes between two interfaces.
Here $k_{1\bot}=\sqrt{\epsilon_1k_0^2-k_\|^2}$, while $
k_{2r\bot}$ and $k_{2l\bot}$ are calculated according to \eref{19}
or \eref{19a} and \eref{21}, respectively.
\begin{figure}[t!]
{\par\centering\resizebox*{8cm}{!}{\includegraphics{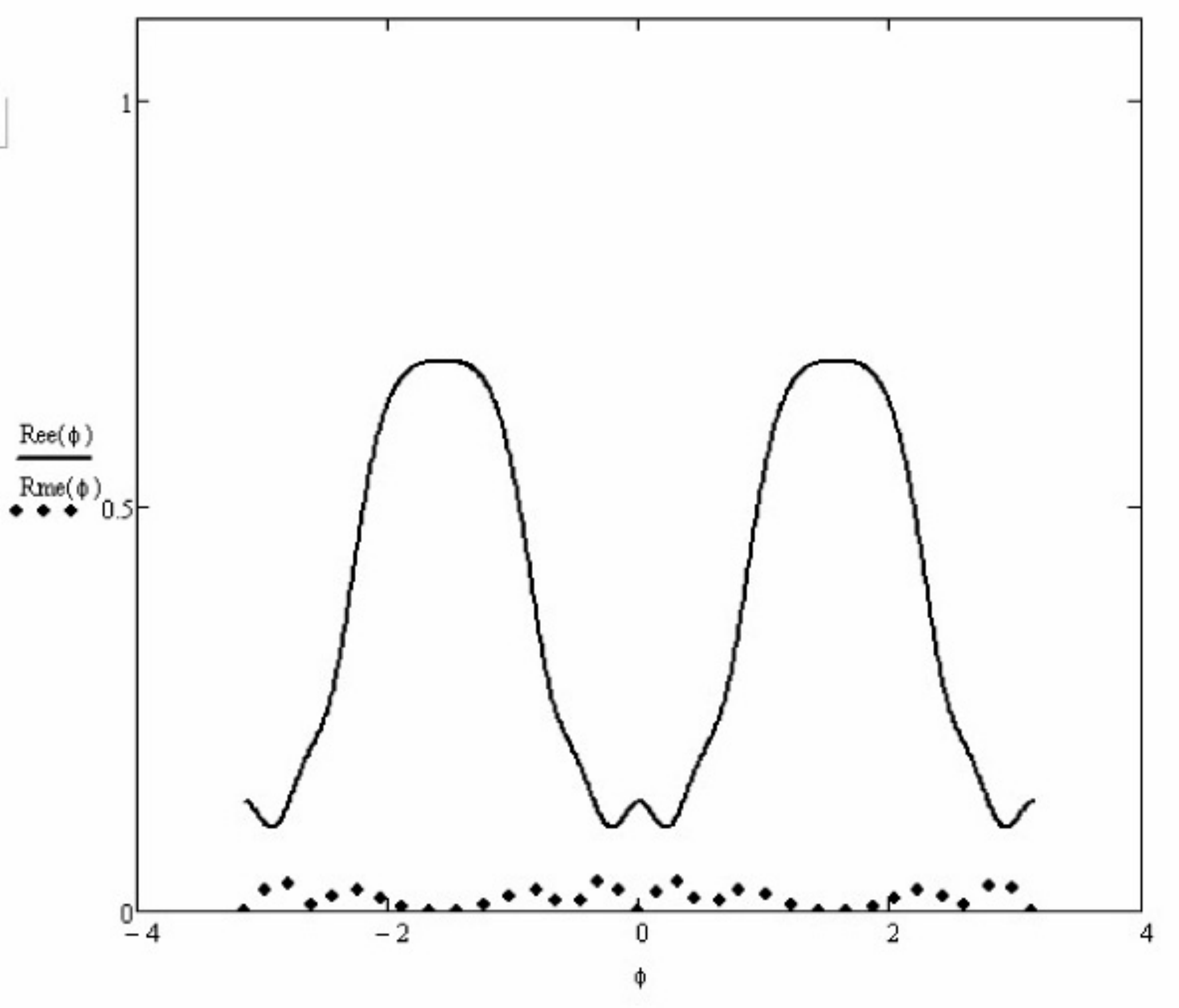}}\par}
\caption{Dependence of reflectivities $|R_{ee}|^2$ and
$|R_{me}|^2$ of an anisotropic plate with $\epsilon_1=1.6$,
$\eta=0.8$ and dimensionless thickness $L\omega/c=10$ on angle
$\phi$ of the plate rotation around its normal, when the
anisotropy vector $\av$ is parallel to interfaces and at $\phi=0$
is directed along $\kk_\|$. The incidence angle $\theta$ is given
by $\sin\theta=0.9$.} \label{r}
\end{figure}

It is very easy to put down a self consistent equation for
determination of $|\ora x\rangle$:
\begin{equation}\label{ksx2}
|\ora x\rangle=\ora\hE(L)\ora\Ta|\xi_0\rangle+\ora\hE(L)
\ola{\Rho'}\ola\hE(L)\ora{\Rho'}|\ora x\rangle.
\end{equation}
The first term at the right hand side describes the incident state
transmitted through the first interface and propagated up to the
second one. The second term describes contribution to the state
$|\ora x\rangle$ of the $|\ora x\rangle$ itself. After reflection
from the second interface this state propagates to the left up to
the first interface, and after reflection from it propagates back
to the point $z=L$. Two terms at the right hand side of \eref{ksx}
add together, which results to some new state. But we denoted it
$|\ora x\rangle$, and it explains derivation of the equation
\eref{ksx}.

From \eref{ksx} we can directly find
\begin{equation}\label{ksx1}
|\ora x\rangle=\Big[\hI-\ora\hE(L)
\ola{\Rho'}\ola\hE(L)\ora{\Rho'}\Big]^{-1}\ora\hE(L)\ora\Ta|\xi_0\rangle,
\end{equation}
and substitution into \eref{ax} and \eref{ksx} gives
\begin{equation}\label{kx2}
\ora\hT(L)\equiv\left(%
\begin{array}{cc}
  T_{ee} & T_{em} \\
  T_{me} & T_{mm} \\
\end{array}%
\right)=\ora{\Ta^{\,\prime}}\Big[\hI-\ora\hE(L)
\ola{\Rho'}\ola\hE(L)\ora{\Rho'}\Big]^{-1}\ora\hE(L)\ora\Ta,
\end{equation}
\begin{equation}\label{kx3}
\ora\hR(L)\equiv\left(%
\begin{array}{cc}
  R_{ee} & R_{em} \\
  R_{me} & R_{mm} \\
\end{array}%
\right)=\ora\Rho+\ola{\Ta^{\,\prime}}\ola\hE(L)\ora{\Rho'}\Big[\hI-\ora\hE(L)
\ola{\Rho'}\ola\hE(L)\ora{\Rho'}\Big]^{-1}\ora\hE(L)\ora\Ta.
\end{equation}

\begin{figure}[t!]
{\par\centering\resizebox*{8cm}{!}{\includegraphics{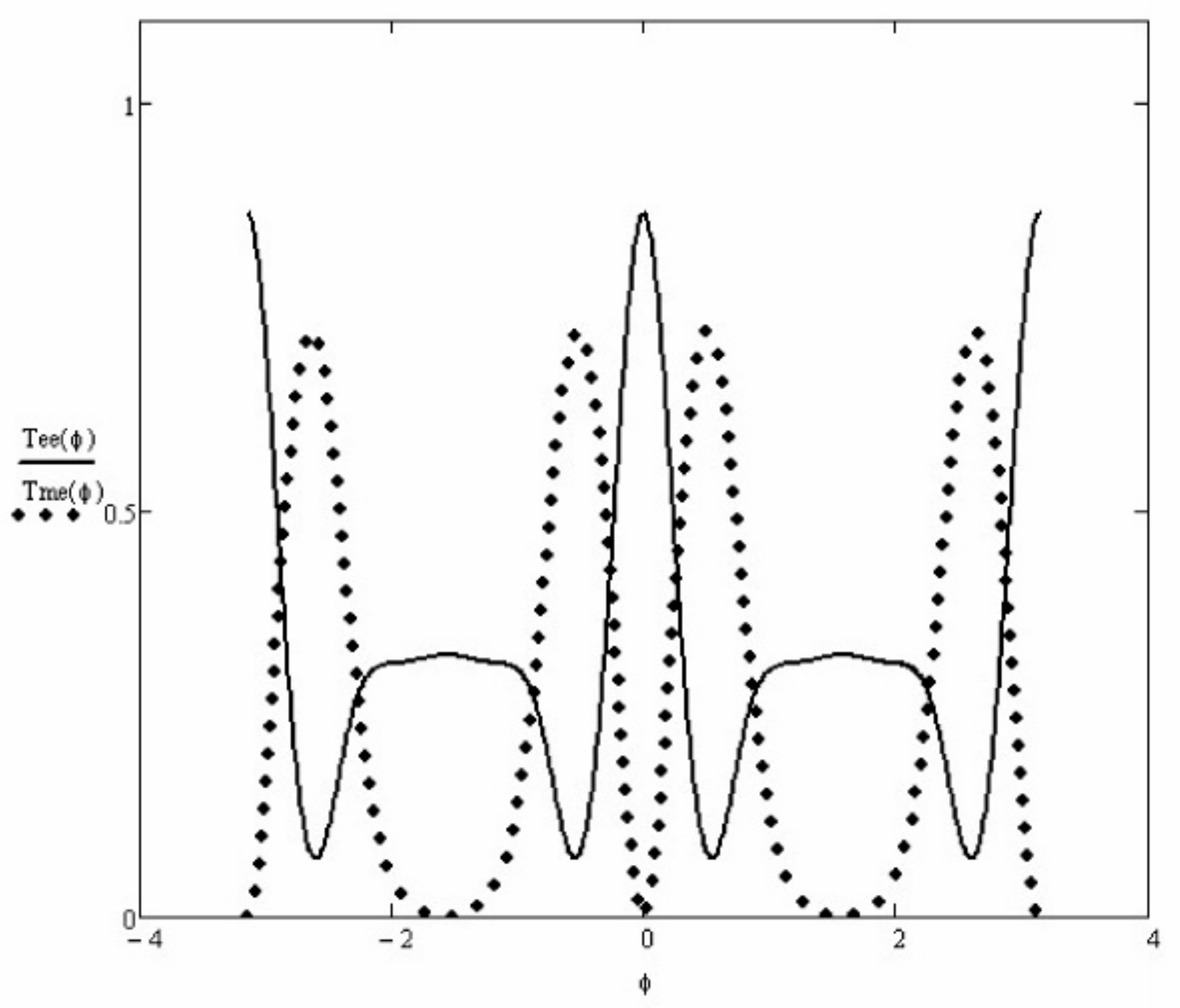}}\par}
\caption{Dependence of transmissivities $|T_{ee}|^2$ and
$|T_{me}|^2$ of an anisotropic plate on angle $\phi$ of the plate
rotation around its normal, with all the parameters the same as
shown in caption of fig.\ref{r}.} \label{t}
\end{figure}

With these formulas we can easily calculate all the reflectivities
and transmissivities
\begin{equation}\label{kx3x}
\left(%
\begin{array}{cc}
  |R_{ee}|^2 & |R_{em}|^2 \\
  |R_{me}|^2 & |R_{mm}|^2\\
\end{array}%
\right), \qquad
\left(%
\begin{array}{cc}
  |T_{ee}|^2 & |T_{em}|^2 \\
  |T_{me}|^2 & |T_{mm}|^2 \\
\end{array}%
\right)
\end{equation}
for arbitrary parameters, arbitrary incidence angles, arbitrary
incident polarizations and arbitrary direction of the anisotropy
vector $\av$. In fig.\ref{r} we present, for example,
reflectivities of TE-mode wave from a plate of thickness $L$ such,
that $L\omega/c=10$. The anisotropy vector is parallel to
interfaces. Therefore, its orientation with respect to wave vector
$\kk_0$ of the incident wave varies with rotation of the plate by
an angle $\phi$ around its normal. The transmissivities of the
same plate in dependence on the angle $\phi$ are presented in
fig.\ref{t}.

\section{D'yakonov surface waves (Dsw)}

Above we found that a surface wave can appear on the interface at
total reflection of a mixed mode at some incident angles. This
surface wave, however, is tied to the incident and reflected wave
and does not exist without them. D'yakonov in 1988 had
discovered~\cite{dya} (see also [18-21]) that on the surface of a
uniaxial anisotropic medium there can exist free surface waves,
analogous to elastic Rayleigh waves on a free surface. We will
derive them with our tensor \eref{1} and a little bit correct
previous derivation by D'yakonov~\cite{dya}.

Let's again consider the space separated by a plane at $z=0$ to
two halves, as shown in fig.\ref{f2}. The left part ($z<0$)
corresponds to anisotropic medium with dielectric permittivity
\eref{1} and the right part is an isotropic medium with dielectric
permittivity $\epsilon_i$.

The surface wave is characterized by the wave function
\begin{equation}\label{dy1}
\Ps(\rr,t)=\lt[\Theta(z<0)\lt(\ps_1\exp(p_lz)+\ps_2\exp(p_2z)\rt)+\Theta(z>0)\ps_i\exp(-p_iz)\rt]e^{i\skk_\|\srr_\|-i\omega
t},
\end{equation}
where $\ps=\cE+\cH$,
\begin{equation}\label{dy}
\cH_{1,2}=\fr{k_\|}{k_0}\lt([\lv\times\cE_{1,2}]-iq_{1,2}[\n\times\cE_{1,2}]\rt),\qquad
\cH_i=\fr{k_\|}{k_0}\lt([\lv\times\cE_i]+iq_i[\n\times\cE_i]\rt),
\end{equation}
parameters $p_{1,2,i}$ provide exponential decay of the surface
wave away from the interface. In \eref{dy} we also introduced
dimensionless parameters $q_{1,2,i}=p_{1,2,i}/k_\|$. For
$\cE_{1,i}$ we have
\begin{equation}\label{dya}
q_{1,i}=\sqrt{1-\epsilon_{1,i}z},
\end{equation}
respectively, where we denoted $z=k_0^2/k_\|^2$. Parameters
$q_{1,i}$ are positive reals when $\epsilon_{1,i}z<1$.

To find $q_2$ for the field $\cE_2$ we need to solve the equation
\begin{equation}\label{a18}
1-x^2+\eta((\lv\cdot\av)-ix(\n\cdot\av))^2=z\epsilon_1(1+\eta),
\end{equation}
where $x$ denotes $q_2$, $\n$ is a unit vector of normal, directed
toward isotropic medium, and $\lv$ is a unit vector along
$\kk_\|$, which together with $\n$ constitutes the plane of
incidence. From this equation it is seen, that $q_2$ can be real
only if vector $\av$ is perpendicular to $\n$ or to $\lv$. In the
first case the axis of anisotropy is parallel to the
interface~\cite{dya}:
\begin{equation}\label{av}
\av=a_l\lv+a_t\t=\cos\theta\lv+\sin\theta\t,
\end{equation}
and solution of \eref{a18} is
\begin{equation}\label{a19}
q_2=\sqrt{1+\eta\cos^2\theta-\epsilon_1z(1+\eta)}.
\end{equation}
It is seen that $q_2(\theta)$ is positive real when
\begin{equation}\label{a199}
z\epsilon_1\fr{1+\eta}{1+\eta\cos^2\theta}\equiv
z\epsilon_2(\theta)<1.
\end{equation}

In the second case
\begin{equation}\label{1av}
\av=a_n\n+a_t\t=\cos\phi\n+\sin\phi\t,
\end{equation}
and solution of \eref{a18} is
\begin{equation}\label{a19a}
q_2(\phi)=\fr{\sqrt{1-\epsilon_1(1+\eta)z}}{\sqrt{1+\eta\cos^2\phi}}.
\end{equation}
Below we will show that in the second case free surface waves do
not exist.

\subsection{Anisotropy axis is parallel to the interface}

When $\av$ is parallel to the interface, vectors $\cE_{1,2}$,
according to section \eref{se1}, can be represented as
\begin{equation}\label{dy2}
\cE_1=-\fr{C_1}{k_\|}[\av\times\kk]=C_1\lt(\sin\theta\n+iq_1[\av\times\n]\rt)=C_1\lt(\sin\theta\n+iq_1\sin\theta\lv-iq_1\cos\theta\t\rt),
\end{equation}
\begin{equation}\label{a12}
\cE_2=\tilde
C_2\lt[\av-\fr{\kk(\av\kk)}{k_\|^2}\fr{1+\eta}{1-q_2^2+\eta(\lv\cdot\av)^2}\rt]=\tilde
C_2\lt[\av- (\lv-iq_2\n)(\av\lv)\fr{1}{1-q_1^2}\rt]=$$
$$=C_2\lt(iq_2\cos\theta\n-q_1^2\cos\theta\lv+
(1-q_1^2)\sin\theta\t\rt),
\end{equation}
where $C_2=\tilde C_2/(1-q_1^2)$, $C_{1,2}$ are some complex
coefficients, and we used relation \eref{a19}.

In the basis $\n$, $\lv$, $\t$, shown in fig.\ref{f2},
polarization $\ora\cE_i$ in isotropic medium can be represented as
\begin{equation}\label{dy3}
\cE_i=\alpha\n+\beta\lv+\gamma\t
\end{equation}
with coordinates $\alpha$, $\beta$ and $\gamma$. Because of
equation $\epsilon_i\na\cdot\cE_i=0$, which is equivalent to
\begin{equation}\label{dy4}
iq_i\alpha+\beta=0,
\end{equation}
vector \eref{dy3} is reduced to
\begin{equation}\label{dy5}
\cE_i=\alpha(\n-iq_i\lv)+\gamma\t.
\end{equation}

From continuity of $\t$- and $\lv$-components of electric field at
the interface we obtain two equations
\begin{equation}\label{dy6}
iC_1q_1\sin\theta-C_2q_1^2\cos\theta=-iq_i\alpha,$$
$$-iC_1q_1\cos\theta+C_2(1-q_1^2)\sin\theta=\gamma.
\end{equation}
To get another two equations we need to use continuity of
tangential components of magnetic fields. Substituting \eref{dy2},
\eref{a12} and \eref{dy5} into \eref{dy} and neglecting common
factor $k_\|/k_0$ we obtain
\begin{equation}\label{ady}
\cH_{1}=C_1\lt(-iq_1\cos\theta\n+q_1^2\cos\theta\lv-[1-q_1^2]\sin\theta\t\rt),$$
$$\cH_{2}=C_2(1-q_1^2)\lt(\sin\theta\n+iq_{2}\sin\theta\lv-iq_2\cos\theta\t\rt),$$
$$\cH_i=\lt(\gamma\n-iq_i\gamma\lv-\alpha(1-q_i^2)\t\rt).
\end{equation}
Continuity of $\lv$ and $\t$ components gives
\begin{equation}\label{bdy}
q_1^2C_1\cos\theta +iC_2q_{2}(1-q_1^2)\sin\theta =-iq_i\gamma,
\end{equation}
\begin{equation}\label{cdy}
C_1\sin\theta +iq_2C_2\cos\theta =\alpha\varepsilon,
\end{equation}
where we denoted $\varepsilon=\epsilon_i/\epsilon_1$. If we
exclude $\gamma$ and $\alpha$ from these equations, we obtain a
homogeneous system of 2 equations for $C_{1,2}$
\begin{equation}\label{dy1y}
q_1\cos\theta(q_i+q_1)C_1+
iC_2(q_{2}+q_i)\epsilon_1z\sin\theta=0,$$
$$
iC_1\sin\theta[q_1\varepsilon+q_i]-C_2\cos\theta\lt(\varepsilon
q_1^2+q_iq_2\rt) =0.
\end{equation}
The system of the two linear equations \eref{dy1y} has solution
only if the determinant of its coefficients is equal to zero,
which gives an equation for $z=k_0^2/k_\|^2$:
\begin{equation}\label{edy}
f(z) =q_1(q_1+q_i)\lt(\varepsilon
q_1^2+q_iq_2\rt)\cos^2\theta-\epsilon_1z(q_{2}+q_i)[q_1\varepsilon+q_i]\sin^2\theta=0.
\end{equation}
Solution of this equation gives the speed of the Dsw
$c_D(z)=c\sqrt z$.

\begin{figure}[t!]
{\par\centering\resizebox*{8cm}{!}{\includegraphics{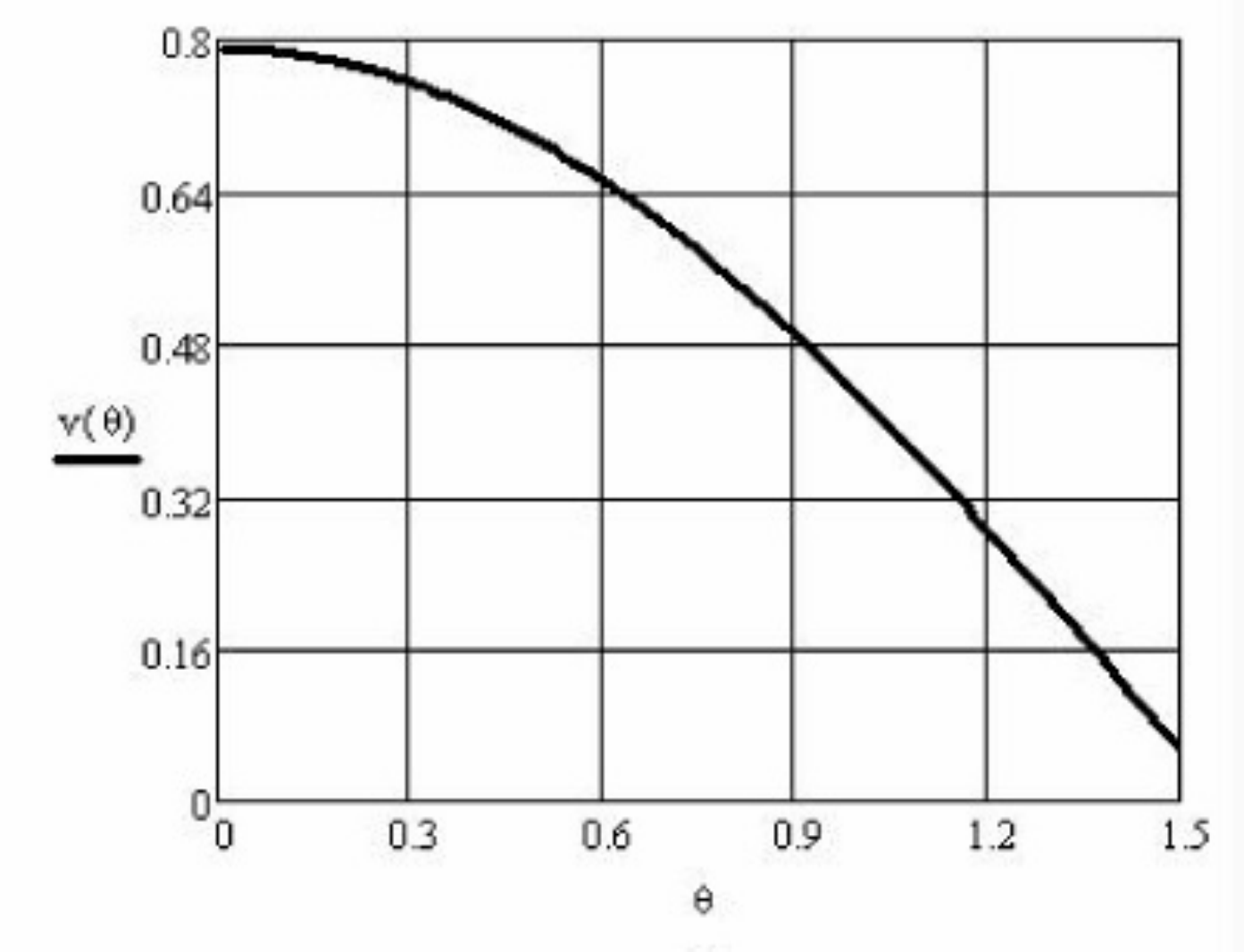}}\par}
\caption{\label{f7} Dependence $v(\theta)=C_D(x)/c$ on angle
$\theta$ between anisotropy axis $\av$ and direction $\kk_\|$ of
the surface wave propagation. The curve was calculated for
$\epsilon_1=1.6$, $\eta=0.4$ and $\epsilon_i=1$.}
\end{figure}

We derived equation \eref{edy} so scrupulously, to show that the
result \eref{edy} slightly differs from the one presented by
equation (8) of~\cite{dya}, and it is not reducible to Eq.(9)
of~\cite{dya}, because solution of \eref{edy} exists even for
$\epsilon_i<\epsilon_1$ ($\epsilon_1$ is denoted $\epsilon_\bot$
in~\cite{dya}). More over it follows from \eref{edy} that the
surface wave exists in much larger range of angles $\theta$ than
was obtained in~\cite{dya}. For instance, in fig. \ref{f7} it is
shown the dependence of ratio $v(\theta)=c_D(z)/c$ on $\theta$, it
is seen that solution of \eref{edy} exists in the full range
$0<\theta<\pi/2$ for $\epsilon_1=1.6$, $\eta=0.4$ and
$\epsilon_i=1$.

\subsection{Anisotropy axis is perpendicular to the propagation direction}

When $\av\bot\lv$ then vectors $\cE_{1,2}$, according to section
\eref{se1}, and $\cH_{1,2}$ according to \eref{dy} can be
represented as
\begin{equation}\label{dyy2}
\cE_1=-\fr{C_1}{k_\|}[\av\times\kk]=C_1\lt(\sin\phi\n+iq_1\sin\phi\lv-\cos\phi\t\rt),
\end{equation}
\begin{equation}\label{xa12}
\cE_2=C_2\lt(\lt[1-q_1^2+q_2^2\rt]\cos\phi\n+iq_2\cos\phi\lv+
\sin\theta\t\rt),
\end{equation}
\begin{equation}\label{ady9}
\cH_{1}=C_1\lt(-\cos\phi\n-iq_1\cos\phi\lv-[1-q_1^2]\sin\phi\t\rt),$$
$$\cH_{2}=C_2(1-q_1^2)\lt(\sin\phi\n+iq_{2}\sin\phi\lv-\cos\phi\t\rt),
\end{equation}
and $\cE_{i}$, $\cH_{i}$ are the same as \eref{dy5} and in
\eref{ady} respectively. After performing the same procedure as
above we obtain an equation for $z=k_0^2/k_\|^2$ in the form
\begin{equation}\label{edyy}
f_1(z)=\lt(\fr{\epsilon_i}{\epsilon_1}q_2+q_i\rt)\cos^2\phi
+(q_{2}+q_i)(1-q_1q_i)\sin^2\phi=0,
\end{equation}
which has no solution because all the terms in it are positive.
Therefore the surface waves do not exist at such orientations of
axis $\av$, as was correctly pointed out in~\cite{dya}.

\section{Conclusion-summary}

\hfill\parbox{9.3cm}{{\it --- I have no confidence that these
authors can form professionally written, carefully researched work
with a decent literature search.}\\\small Referee of Am.J.Phys.}
\medskip

In the case of uniaxial or biaxial anisotropic media we used for
the tensor of dielectric permittivity $\epsilon_{ij}$ in the form
\begin{equation}\label{c-s1}
\eps_{ij}=\epsilon_1\lt[\delta_{ij}+\eta a_ia_j\rt],\qquad
\eps_{ij}=\epsilon_1\lt[\delta_{ij}+\eta_aa_ia_j+\eta_bb_ib_j\rt],
\end{equation}
where $\epsilon_1$ is a parameter of isotropic part of the
tensors, $\av$, $\bb$ are the unit vectors along axes of
anisotropy, and $\eta$, $\eta_{a,b}$ are respective anisotropy
parameters. With such tensors we can immediately find for a plain
wave $\cE\exp(i\kk\rr-i\omega t)$ with an arbitrary direction
$\ka=\kk/k$ of propagation analytical expressions for the
polarization vector $\cE$ and wave number $k(\omega)$. In the case
of uniaxial anisotropic medium we found that only two modes of
linear polarizations can propagate inside it. One is transverse
mode with
\begin{equation}\label{c-s2}
\cE_1=[\av\times\ka],\qquad k_1=(\omega/c)\sqrt{\epsilon_1},
\end{equation}
and another one is the mixed mode (it has a component of
polarization parallel to the wave vector)
\begin{equation}\label{c-s3}
\cE_2=\av-\ka(\ka\cdot\av)\fr{\epsilon_2(\theta)}{\epsilon_1},
\qquad k_2=(\omega/c)\sqrt{\epsilon_2(\theta)},\qquad
\epsilon_2(\theta)=\fr{\epsilon_1(1+\eta)}{1+\eta\cos^2\theta},\qquad
\cos\theta=(\av\cdot\ka).
\end{equation}

Next we considered reflection of obtained plain waves from an
interface with an isotropic medium and had shown that reflection
of every mode is accompanied by beam splitting, that the wave of
mode 2 is in general reflected nonspecularly, and at some incident
angles reflection of mode 2 can create a surface wave, which is
bound to the incident and reflected waves of mode 2. The beam
splitting at reflection can be spectacularly demonstrated with the
help of light transmission through an anisotropic cone.

After calculation of reflection from interfaces from inside and
outside anisotropic medium we had shown an algorithm to calculate
reflection and transmission of plane parallel plates without
matching of the wave field at two interfaces. In the case of
normal incidence on the plate of a plane wave with linear
polarization transmitted wave in general has elliptical
polarization. The form of ellipse changes with rotation of the
birefringent plate around its normal and at two distinct
orthogonal direction the ellipse reduces to linear polarization
identical to that of the incident wave.

Next we considered the free Dsw on the surface of anisotropic
media. We corrected some error in derivation presented
in~\cite{dya} and have shown that Dsw exist in larger range of
variation of dielectric constants and in a larger range of angles
between direction of surface wave propagation and direction of
anisotropy vector $\av$.

\begin{figure}[t!]
{\par\centering\resizebox*{8cm}{!}{\includegraphics{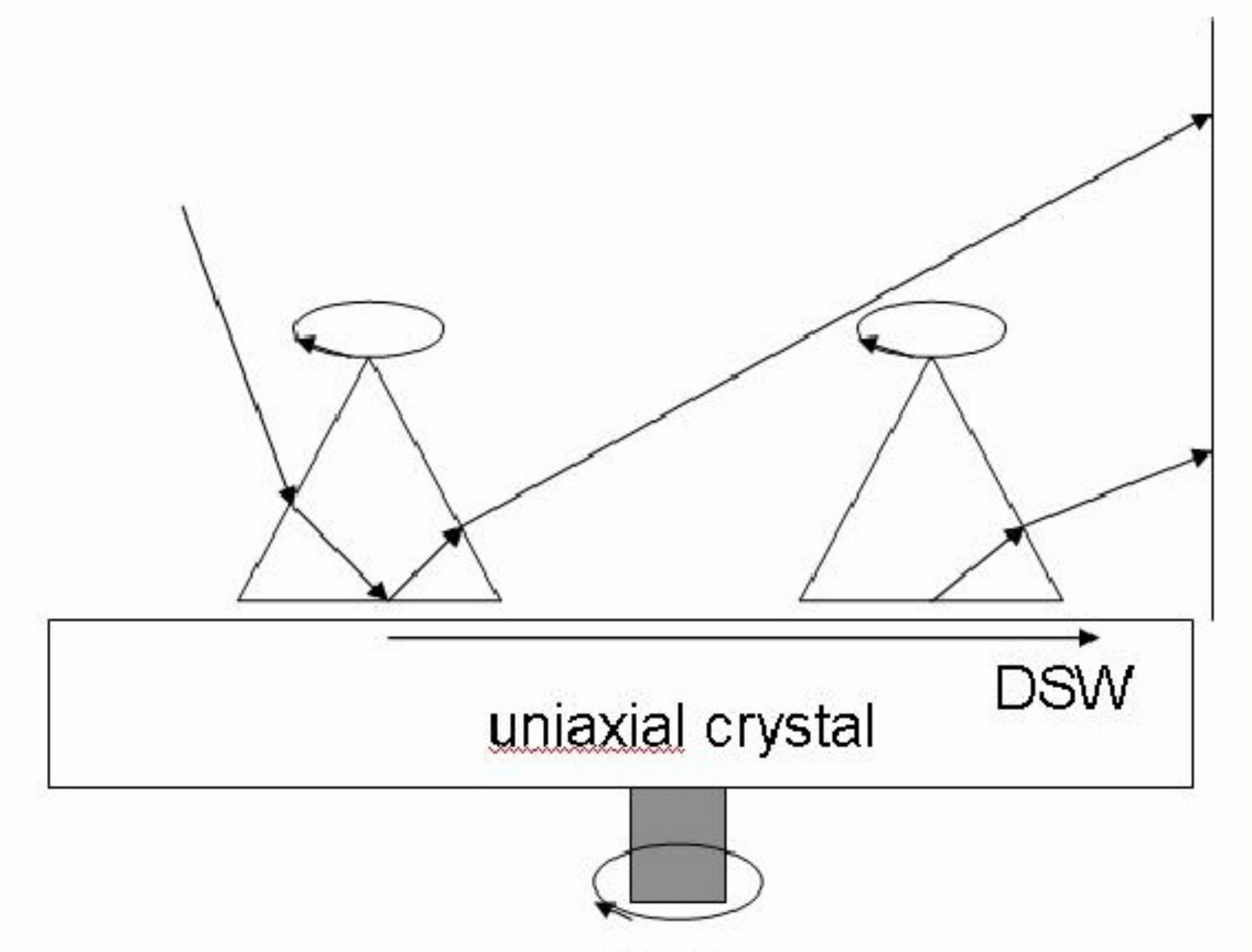}}\par}
\caption{\label{f8} Scheme of a possible experiment to observe
Dsw.}
\end{figure}

To observe the D'yakonov surface waves it is possible to use the
experimental scheme shown in fig. \ref{f8}, which is different
comparing to the one used in~\cite{dya5}. Disc of a uniaxial
crystal with anisotropy axis $\av$ parallel to the surface can be
pivoted around its axis to change angle between direction of Dsw
propagation $\ka=\kk_\|/k_\|$ and the vector $\av$. The Dsw is
excited at frustrated total reflection in an anisotropic cone
similar to that one shown in fig. \ref{f} (here for simplicity we
draw only one transmitted ray). Excitation takes place only when
speed of the incident or reflected wave inside cone matches the
speed of Dsw. Rotation of the cone around its axis permits some
tuning of the speed.

The second anisotropic cone identical to the first one detects
Dsw, and the light transmitted into it through the small gap
should be visible on a screen, as shown in fig. \ref{f8}.

\section*{Acknowledgement}
We are grateful to Prof. Lukas Novotny, Dr. Svetlana Lukishova and
Dr. Sergio G. Rodrigo from Institute of optics of Rochester
University of N.Y. USA for their assistance.

\appendix
\section{Waves in a two axes anisotropic medium}

A two-axial anisotropic medium is characterized by two unit
vectors $\av$ and $\bb$, and two anisotropy parameters
$\epsilon'_a$ and $\epsilon'_a$. Therefore the tensor $\eps$ has
matrix elements
\begin{equation}\label{1a}
\eps_{ij}=\epsilon_1\delta_{ij}+\epsilon'_aa_ia_j+\epsilon'_bb_ib_j,
\end{equation}
and equations \eref{a4}, \eref{6} and  \eref{b4} take the
respective forms
\begin{equation}\label{1b4}
\eps\cE=\epsilon_1\cE+\epsilon'_a\av(\av\cdot\cE)+\epsilon'_b\bb(\bb\cdot\cE),
\end{equation}
\begin{equation}\label{6a1}
(\ka\cdot\cE)+\eta_a(\ka\cdot\av)(\av\cdot\cE)+\eta_b(\ka\cdot\bb)(\bb\cdot\cE)=0.
\end{equation}
\begin{equation}\label{4a1}
(k^2-k_0^2\epsilon_1)\cE-k^2\ka(\ka\cdot\cE)-k_0^2\epsilon_1\eta_a\av(\av\cdot\cE)-k_0^2\epsilon_1\eta_b\bb(\bb\cdot\cE)=0.
\end{equation}
In the last two equations we introduced notations
$\eta_a=\epsilon'_a/\epsilon_1$, and
$\eta_b=\epsilon'_b/\epsilon_1$. For simplicity, we assume that
$\av\bot\bb$, introduce the orthonormal basis $\av$, $\bb$,
$\cc=[\av\times\bb]$ and in this basis represent
\begin{equation}\label{66}
\cE=\alpha\av+\beta\bb+\gamma\cc
\end{equation}
with coordinates $\alpha$, $\beta$, $\gamma$, which are not
completely independent, because \eref{66} should satisfy
\eref{6a1}. Substituting \eref{66} into \eref{6a1} gives
\begin{equation}\label{15}
\alpha(\kk\cdot\av)+\beta(\kk\cdot\bb)
+\gamma(\kk\cdot\cc)]+\eta_a(\kk\cdot\av)\alpha+
\eta_b(\kk\cdot\bb)\beta=0.
\end{equation}
Therefore
\begin{equation}\label{6a2}
\gamma(\ka\cdot\cc)=-\alpha(\ka\cdot\av)(1+\eta_a)
-\beta(\ka\cdot\bb)(1+\eta_b).
\end{equation}
From \eref{66} we also obtain that
\begin{equation}\label{66a}
(\cE\cdot\av)=\alpha, \qquad (\cE\cdot\bb)=\beta.
\end{equation}
Now let's substitute \eref{66a} and $(\ka\cdot\cE)$ from
\eref{6a1} into \eref{4a1} and multiply \eref{4a1} consecutively
by $\av$ and $\bb$. As a result we obtain a system of two linear
equations
\begin{eqnarray}\label{66b}
  \Big(k^2[1+\eta_a(\ka\cdot\av)^2]-k^2_0\epsilon_1(1+\eta_a)\Big)\alpha+\eta_bk^2(\ka\cdot\av)(\ka\cdot\bb)\beta &=& 0
  \\\nonumber
  \Big(k^2[1+\eta_b(\ka\cdot\bb)^2]-k^2_0\epsilon_1(1+\eta_b)\Big)\beta+\eta_ak^2(\ka\cdot\av)(\ka\cdot\bb)\alpha &=&
  0.
\end{eqnarray}
The solution to this system exists if the determinant is equal to
zero. This condition can be written as
\begin{equation}\label{66c}
(k^2-\epsilon_a(\theta_a)k_0^2)(k^2-\epsilon_b(\theta_b)k_0^2)=\fr{\eta_a\eta_bk^2(\ka\cdot\av)^2(\ka\cdot\bb)^2}
{[1+\eta_a(\ka\cdot\av)^2][1+\eta_b(\ka\cdot\bb)^2]}.
\end{equation}
where
\begin{equation}\label{66d}
\epsilon_{a,b}(\theta_{a,b})=\fr{\epsilon_1(1+\eta_{a,b})}{1+\eta_{a,b}\cos^2\theta_{a,b}},\qquad\cos\theta_{a}=(\ka\cdot\av),
\qquad\cos\theta_{b}=(\ka\cdot\bb).
\end{equation}

The solution to \eref{66c} provides two different values of
$k_{1,2}$, for which we find $\alpha$, $\beta$. After substituting
the latter coordinates into \eref{6a2}, we obtain the last
coordinate $\gamma$. Thus we find two different plain waves with
wave vectors $\kk_{1,2}=k_{1,2}\ka$ and linear polarizations
$\cE_{1,2}$ \eref{66}.

\section{Reflection from an interface from inside of the anisotropic medium}

Exclusion of $\ora\tau_{ej}$ from \eref{a23} and \eref{a23a}, and
exclusion of $\ora\tau_{mj}$ from \eref{c23} and \eref{d23a} gives
two equations for $\ora\rho_{1j}$, $\ora\rho_{2j}$, which is
convenient to represent in the matrix form {\small
\begin{equation}\label{m}
\left(%
\begin{array}{cc}
 \lt((\lv\cdot\ola\cH_1)+\kappa_{0\bot}(\t\cdot\ola\cE_1)\rt) &
 \lt((\lv\cdot\ola\cH_2)+\kappa_{0\bot}(\t\cdot\ola\cE_2)\rt) \\
\lt(\kappa_{0\bot}(\t\cdot\ola\cH_1)-(\lv\cdot\ola\cE_1)\rt) &
\lt(\kappa_{0\bot}(\t\cdot\ola\cH_2)-(\lv\cdot\ola\cE_2)\rt) \\
\end{array}%
\right){\ora\rho_{1j}\choose\ora\rho_{2j}}=-{(\lv\cdot\ora\cH_j)+\kappa_{0\bot}(\t\cdot\ora\cE_j)\choose
\kappa_{0\bot}(\t\cdot\ora\cH_j)-(\lv\cdot\ora\cE_j)}.
\end{equation}}
Solution of this equation is {\small
\begin{equation}\label{mm}
{\ora\rho_{1j}\choose\ora\rho_{2j}}=\fr{-1}D\left(%
\begin{array}{cc}
 \lt(\kappa_{0\bot}(\t\cdot\ola\cH_2)-(\lv\cdot\ola\cE_2)\rt)&
 -\lt((\lv\cdot\ola\cH_2)+\kappa_{0\bot}(\t\cdot\ola\cE_2)\rt) \\
-\lt(\kappa_{0\bot}(\t\cdot\ola\cH_1)-(\lv\cdot\ola\cE_1)\rt) &
\lt((\lv\cdot\ola\cH_1)+\kappa_{0\bot}(\t\cdot\ola\cE_1)\rt)\\
\end{array}%
\right){(\lv\cdot\ora\cH_j)+\kappa_{0\bot}(\t\cdot\ora\cE_j)\choose
\kappa_{0\bot}(\t\cdot\ora\cH_j)-(\lv\cdot\ora\cE_j)},
\end{equation}}
Where $D$ is determinant
$$D=\lt((\lv\cdot\ola\cH_1)+\kappa_{0\bot}(\t\cdot\ola\cE_1)\rt)\lt(\kappa_{0\bot}
(\t\cdot\ola\cH_2)-(\lv\cdot\ola\cE_2)\rt)-$$
\begin{equation}\label{dt}
-\lt((\lv\cdot\ola\cH_2)+\kappa_{0\bot}(\t\cdot\ola\cE_2)\rt)\lt(\kappa_{0\bot}(\t\cdot\ola\cH_1)-(\lv\cdot\ola\cE_1)\rt).
\end{equation}
Substitution of these expressions into \eref{a23} and \eref{c23}
gives refraction amplitudes $\ora\tau_{e,mj}$
\begin{equation}\label{mm1}
{\ora\tau_{ej}\choose\ora\tau_{mj}}={(\t\cdot\ora\cE_j)\choose(\t\cdot\ora\cH_j)}+\left(%
\begin{array}{cc}
 (\t\cdot\ola\cE_1)&
 (\t\cdot\ola\cE_2) \\
(\t\cdot\ola\cH_1) &
(\t\cdot\ola\cH_2)\\
\end{array}%
\right){\ora\rho_{1j}\choose\ora\rho_{2j}},
\end{equation}

\subsubsection{The most general case}

Above we considered the case when the incident wave has
polarization vector $\e_{j}$ with unit amplitude. (We remind that
vectors $\e_{j}$ can be not normalized to unity.) To find later
reflections from plane parallel plates we will need a more general
case, when the incident wave has both modes with amplitudes
$x_{1,2}$. To find amplitudes of reflected and transmitted waves
in the general case it is convenient to represent the state of the
incident wave in the form of 2 dimensional vector
\begin{equation}\label{mx}
|\ora x\rangle={\ora x_1\choose \ora x_2}.
\end{equation}
then the states of reflected and refracted waves are also
described by 2-dimensional vectors, which can be represented as
\begin{equation}\label{m10}
|\ola\psi\rangle={\ola \psi_1\choose \ola \psi_2}=\ora{\Rho'}|\ora
x\rangle,\qquad |\ora\psi_0\rangle={\ora \psi_e\choose \ora
\psi_m}=\ora{\Ta^{\,\prime}}|\ora x\rangle,
\end{equation}
where $\ora{\Rho'}$ and $\ora{\Ta^{\,\prime}}$ are two dimensional
matrices
\begin{equation}\label{xa3a}
\ora{\Rho'}=\left(%
\begin{array}{cc}
  \ora\rho_{11} & \ora\rho_{12} \\
  \ora\rho_{21} & \ora\rho_{22} \\
\end{array}%
\right),\qquad \ora{\Ta^{\,\prime}}=\left(%
\begin{array}{cc}
  \ora\tau_{e1} & \ora\tau_{e2} \\
  \ora\tau_{m1} & \ora\tau_{m2} \\
\end{array}%
\right).
\end{equation}
We introduced the prime here and below to distinguish refraction
and reflection from inside the medium and the similar matrices
obtained for incident waves outside the medium.

These formulas will be used later for calculation of reflection
and transmission of plane parallel anisotropic plates. In the case
of a plate we have two interfaces, therefore we need also
reflection and refraction at the left interface from inside the
plate. They can be easily found from symmetry considerations.
Their representation is obtained from \eref{mm} --- \eref{mm1} by
reverse of arrows and change of the sign before $\kappa_{0\bot}$.
After this action we find
\begin{equation}\label{x1a3a}
\ola{\Rho'}=\left(%
\begin{array}{cc}
  \ola\rho_{11} & \ola\rho_{12} \\
  \ola\rho_{21} & \ola\rho_{22} \\
\end{array}%
\right),\qquad \ola{\Ta^{\,\prime}}=\left(%
\begin{array}{cc}
  \ola\tau_{e1} & \ola\tau_{e2} \\
  \ola\tau_{m1} & \ola\tau_{m2} \\
\end{array}%
\right).
\end{equation}
Reflection from outside the medium is to be considered separately.

\subsubsection{Energy conservation}

It is always necessary to control correctness of the obtained
formulas. One of the best controls is the test of energy
conservation. One should always check whether the energy density
flux of incident wave along the normal to interface is equal to
the sum of energy density fluxes of reflected and refracted waves,
and the most important in such tests is the correct definition of
the energy fluxes. In isotropic media it is possible to define
energy flux along a vector $\n$ as
\begin{equation}\label{jn}
(\J\cdot\n)=\fr{(\kk\cdot\n)}k\fr{c}{\sqrt{\epsilon}}\fr{\epsilon
\cE^2+\cH^2}{8\pi},
\end{equation}
or
\begin{equation}\label{jn1}
(\J\cdot\n)=c\fr{(\n\cdot[\cE\times\cH])}{4\pi}.
\end{equation}
In isotropic media both definitions are equivalent, because
$\cH=[\kk\times\cE]/k_0$, and $(\kk\cdot\cE)=0$. The first
definition looks even more preferable since the second one can be
written even for stationary fields, where there are no energy
flux.

In anisotropic media only the second definition is valid, and
because in mode 2 the field $\cE$ is not orthogonal to $\kk$, the
direction of the energy density flux is determined not only by
wave vector, but also by direction of the field $\cE$ itself.

\section{Reflection from an interface from outside of the anisotropic medium}

Exclusion of $\ora\rho_{ee}$ and $\ora\rho_{me}$ leads to
\begin{equation}\label{ee}
\left(%
\begin{array}{cc}
 \lt(\kappa_{0\bot}(\t\ora\cE_{1})-(\lv\ora\cH_{1})\rt) & \lt(\kappa_{0\bot}(\t\ora\cE_{2})-
(\lv\ora\cH_{2})\rt) \\
 \lt(\kappa_{0\bot}(\t\ora\cH_{1})+(\lv\ora\cE_{1})\rt)  & \lt(\kappa_{0\bot}(\t\ora\cH_{2})+(\lv\ora\cE_{2})\rt) \\
\end{array}%
\right){\ora\tau_{1e}\choose
\ora\tau_{2e}}={2\kappa_{0\bot}\choose 0}
\end{equation}
and the solution
\begin{equation}\label{ee1}
{\ora\tau_{1e}\choose
\ora\tau_{2e}}=\fr1{D_e}\left(\begin{array}{cc}
 \lt(\kappa_{0\bot}(\t\ora\cH_{2})+(\lv\ora\cE_{2})\rt) & -\lt(\kappa_{0\bot}(\t\ora\cE_{2})-
(\lv\ora\cH_{2})\rt) \\
 -\lt(\kappa_{0\bot}(\t\ora\cH_{1})+(\lv\ora\cE_{1})\rt)  &\lt(\kappa_{0\bot}(\t\ora\cE_{1})-(\lv\ora\cH_{1})\rt)
  \\
\end{array}%
\right){2\kappa_{0\bot}\choose 0}
\end{equation}
where $D_e$ is determinant
$$D_e=\lt(\kappa_{0\bot}(\t\ora\cE_{1})-(\lv\ora\cH_{1})\rt)\lt(\kappa_{0\bot}(\t\ora\cH_{2})+(\lv\ora\cE_{2})\rt)-$$
\begin{equation}\label{ee2}
-\lt(\kappa_{0\bot}(\t\ora\cE_{2})-
(\lv\ora\cH_{2})\rt)\lt(\kappa_{0\bot}(\t\ora\cH_{1})+(\lv\ora\cE_{1})\rt).
\end{equation}
Substitution of $\ora\tau_{je}$ into \eref{A23} and \eref{C23}
gives
\begin{equation}\label{ee3}
{\ora\rho_{ee}\choose\ora\rho_{me}}=
\left(%
\begin{array}{cc}
  (\t\ora\cE_{1}) & (\t\ora\cE_{2}) \\
  (\t\ora\cH_{1}) & (\t\ora\cH_{2}) \\
\end{array}%
\right){\ora\tau_{1e}\choose \ora\tau_{2e}}-{1\choose0}.
\end{equation}

In the case of incident TM-mode we have boundary conditions
\begin{equation}\label{A23aa}
(\t\ora\cH_{1})\ora\tau_{1m}+(\t\ora\cH_{2})\ora\tau_{2m}=1+\ora\rho_{mm},
\end{equation}
\begin{equation}\label{A2a3a}
(\lv\ora\cE_{1})\ora\tau_{1m}+(\lv\ora\cE_{2})\ora\tau_{2m}=\kappa_{0\bot}(1-\ora\rho_{mm}),
\end{equation}
\begin{equation}\label{C23a}
(\t\ora\cE_{1})\ora\tau_{1m}+(\t\ora\cE_{2})\ora\tau_{2m}=\ora\rho_{em},
\end{equation}
\begin{equation}\label{d2a23a}
(\lv\ora\cH_{1})\ora\tau_{1m}+(\lv\ora\cH_{2})\ora\tau_{2m}=\kappa_{0\bot}\ora\rho_{em}.
\end{equation}
Exclusion of $\ora\rho_{me}$ and $\ora\rho_{mm}$ leads to
\begin{equation}\label{me}
\left(%
\begin{array}{cc}
 \lt(\kappa_{0\bot}(\t\ora\cH_{1})+(\lv\ora\cE_{1})\rt) &
 \lt(\kappa_{0\bot}(\t\ora\cH_{2})+
(\lv\ora\cE_{2})\rt) \\
 \lt(\kappa_{0\bot}(\t\ora\cE_{1})-(\lv\ora\cH_{1})\rt)  & \lt(\kappa_{0\bot}(\t\ora\cE_{2})-(\lv\ora\cH_{2})\rt) \\
\end{array}%
\right){\ora\tau_{1m}\choose
\ora\tau_{2m}}={2\kappa_{0\bot}\choose 0}.
\end{equation}
Therefore
\begin{equation}\label{me1}
{\ora\tau_{1m}\choose
\ora\tau_{2m}}=\fr1{D_m}\left(\begin{array}{cc}
 \lt(\kappa_{0\bot}(\t\ora\cE_{2})-(\lv\ora\cH_{2})\rt) &
 -\lt(\kappa_{0\bot}(\t\ora\cH_{2})+
(\lv\ora\cE_{2})\rt) \\
 -\lt(\kappa_{0\bot}(\t\ora\cE_{1})-(\lv\ora\cH_{1})\rt)  &\lt(\kappa_{0\bot}(\t\ora\cH_{1})+(\lv\ora\cE_{1})\rt)
  \\
\end{array}%
\right){2\kappa_{0\bot}\choose 0},
\end{equation}
where $D_m=-D_e$ \eref{ee2}. Substitution of $\ora\tau_{jm}$ into
\eref{A23aa} and \eref{C23a} gives
\begin{equation}\label{me2}
{\ora\rho_{em}\choose\ora\rho_{mm}}=\left(%
\begin{array}{cc}
  (\t\ora\cE_{1})& (\t\ora\cE_{2}) \\
  (\t\ora\cH_{1}) & (\t\ora\cH_{2})\\
\end{array}%
\right){\ora\tau_{1m}\choose\ora\tau_{2m}}-{0\choose1}.
\end{equation}
In the general case, when the incident wave has an amplitude
$\ora\xi_e$ in TE-mode and amplitude $\ora\xi_m$ in TM-mode, the
state of the incident wave can be described by two-dimensional
vector
\begin{equation}\label{me3}
|\ora\xi_0\rangle={\ora\xi_e\choose\ora\xi_m},
\end{equation}
and the states of reflected and transmitted waves can be
represented as
\begin{equation}\label{me4}
|\ola\xi_0\rangle={\ola\xi_e\choose \ola\xi_m}=\ora\Rho|\ora
\xi_0\rangle,\qquad |\ora\xi\rangle={\ora \xi_1\choose \ora
\xi_2}=\ora\Ta|\ora\xi_0\rangle,
\end{equation}
where $\ora\Rho$ and $\ora\Ta$ are the two dimensional matrices
\begin{equation}\label{xa3aa}
\ora\Rho=\left(%
\begin{array}{cc}
  \ora\rho_{ee} & \ora\rho_{em} \\
  \ora\rho_{me} & \ora\rho_{mm} \\
\end{array}%
\right),\qquad \ora\Ta=\left(%
\begin{array}{cc}
  \ora\tau_{1e} & \ora\tau_{1m} \\
  \ora\tau_{2e} & \ora\tau_{2m} \\
\end{array}%
\right).
\end{equation}

\end{document}